\documentclass[a4paper,12pt]{article}
\usepackage{amsmath}
\usepackage{amsfonts}
\usepackage{graphicx}   
\usepackage{amssymb}
\usepackage{color}

\title{Three-dimensional Quantum Slit Diffraction and Diffraction in Time.}
\author{M. Beau \& T. C. Dorlas \\Dublin Institute for Advanced Studies \\
School of Theoretical Physics \\ 10 Burlington Road, Dublin 4,
Ireland. }

\begin{document}

\maketitle

\begin{abstract}
We study the quantum slit diffraction problem in three dimensions. 
In the treatment  of diffraction of particles by a slit, 
it is usually assumed that the motion perpendicular to the slit is classical.
Here we take into account the effect of the quantum nature of the motion perpendicular to the slit using 
the Green function approach \cite{Zeilinger}. 
We treat the diffraction of a Gaussian wave packet for general boundary conditions on the shutter.
The difference between the standard and our three-dimensional slit diffraction models
is analogous to the diffraction in time phenomenon introduced in \cite{Moshinsky}.
We derive corrections to the standard formula for the diffraction pattern, 
and we point out situations in which this might be observable.
In particular, we discuss the diffraction in space and time in the presence of gravity. 
\end{abstract}

\newpage

\tableofcontents

\section{Introduction}

The quantum slit diffraction experiment of electrons was first
realized experimentally in 1961 by C. J\"onsson, see
\cite{Jonsson} and \cite{Jonsson2}, but the first experimental
proof of the quantum diffraction for individual electrons was
shown in the seventies by O. Donati, P. G. Merli, G. P. Missiroli
and G. Pozzi, \cite{Pozzi1},\cite{Pozzi2} using electron biprisms
and later independently by A. Tonomura, J. Endo, H. Ezawa, T. Matsuda and T. Kawasaki \cite{Tonomura}.
The quantum diffraction phenomenon has been interpreted via the
famous thought experiment imagined by Richard Feynman in
\cite{Feynman}. We mention that this double slit experiment has
recently also been done experimentally \cite{Bach} in a situation
where the probability distribution for individual electron on the
screen was observed (statistically) while varying the position of
a mask hiding one, two or none of the slits. In addition,
nano-slit electron experiments were recently performed, see for example \cite{Frabboni}. 
Furthermore, slit experiments were carried out with neutrons, see \cite{ZeilingerNeutron} and the references therein,
ultracold atoms \cite{Takuma} and with heavy molecules such as $C_{60}$, see \cite{ZeilingerC60}.
 
In \cite{FH}, Feynman and Hibbs also treated in detail a quantum slit
diffraction model using the path integral formalism to compute the
quantum slit propagator. The model consists of a one-dimensional
slit appearing in the motion of an electron at a time $\tau>t_0=0$
and then removed instantaneously, the electron striking the screen
at a time $t > \tau$. Actually, this means that the motion of the
electron from the source to the screen consists of two independent
motions, the first from the source to the slit and the second
from the slit to the screen. Under this hypothesis, the quantum
propagator for the single-slit system can be written as a product
of the free propagator in the $x$-direction orthogonal to the slit
and the propagator along the one-dimensional slit axis $z$: see
\cite{Barut} and \cite{Beau} for pedagogical presentations of this
model. This so-called ``\textit{truncation assumption}'' or 
``\textit{truncation approximation}''\cite{Zecca} is
convenient and valid under certain conditions. First, we suppose
that the particle passes through the aperture at the classical
time $\tau=t_c=D/v_x=(D/x)t$, where $D$ is the distance between
the emittor and the center of the slit, $x$ is the distance
between the emittor and the screen, and $v_x=x/t$ is the classical
velocity related to the wave length $\lambda$ by the de Broglie
relation $\lambda\approx 2\pi\hbar/(mv_x)= 2\pi\hbar t/(m x)$.
Here we assume $v\approx v_x$ because we have $a\ll x$, where $a$
is the size of the slit, and we take $z\ll x$ where $z$ is the
position of the particle detected on the screen. 
This assumption means that the motion along the $x$-axis is classical whereas the
one parallel to the screen (in the $z$ direction) is quantum. The
main goal of this article is to find the condition justifying
the latter assumption, i.e. the classical behavior of the particle
along the $x$-axis, independently of the fact that we consider the
aperture of the slit to be relatively small, and also to obtain a
correction to the single-slit propagator formula and to analyse the
resulting effect on the probability density function for
different regimes.

At this stage, we should mention that another curious quantum
diffraction phenomenon was imagined in 1952 M. Moshinsky
\cite{Moshinsky}. He showed that for a monochromatic plane wave
moving along a one-dimensional line, if a perfectly absorbing
screen is placed at a fixed position on the axis at the times
$0<t<t_1$ and is removed at the time $t_1$, the probability
density function would be similar to the one observed for the
diffraction in space by a half-plane. By analogy, therefore, we
call this phenomenon \textit{diffraction in time}. It was
first observed experimentally in 1997 by the cold atom team of the
Kastler Brossel Institute in Paris, see \cite{Dalibard}. In the
mean time, the problem of diffraction in both space and time of
monochromatic plane waves for a perfectly reflective slit-screen
was treated in \cite{Zeilinger}. They used the Green's function
method, giving the general solution of the diffusion equation
\cite{Morse} in the half-space delimited by the plane of the slit
with given boundary conditions. 
Another approach was developed in \cite{Goussev} to construct the quantum propagator 
for a purely absorbing screen.  
Also, some recent progress have been made in \cite{Goussev2} where
an exactly solvable model of diffraction in time for a perfectly absorbing shutter was developed.
There are other recent articles  
\cite{deCampo2}, \cite{deCampo3} and review \cite{deCampo1}, on the diffraction in time phenomenon.
Also we refer the reader to the series of articles dealing with diffraction in space and time
for Gaussian wave packets \cite{Kalbermann1}, \cite{Kalbermann2}, \cite{Kalbermann3}
where the autor discusses the effect of the finite width of the wave packet.

In this article, we use the Brukner-Zeilinger approach
\cite{Zeilinger} to construct the three-dimensional quantum slit propagator of a
Gaussian wave packet for general boundary conditions on the slit (absorbing or reflective),
and to derive a semi-classical formula for the propagator taking into 
account the diffraction in time.
In Section II, we introduce the following
model. Consider a particle, modeled by a three-dimensional
Gaussian wave packet of width $\sigma$, which is emitted at the
time $t_0=0$ from the position $x_0<0,\ y_0=0,\ z_0=0$. The
aperture of the slit is closed until the time
$t_1\geq0$ after which it is opened and the wave packet propagates from the
rectangular aperture of the slit (centered at $x=0$) to a screen
(centered at $x>0$) where the position $(y,z)$ of the particle is detected at a time $t>t_1$. 
In Section III, we derive an explicit integral formula for the single-slit propagator (for a rectangular
aperture) for arbitrary boundary conditions on the plane of the slit, in
the case where the times of emission of the particle and of
opening of the slit coincide: $t_1=t_0=0$. After that, we will
show that there is a \textit{semi-classical transition}, when the
parameter $\mu=m|\bold{r}|^2/(\hbar t)$ is large, where $\bold{r}$
is the position of the particle detected on the screen at the time
$t$. We will also interpret the semi-classical propagator formula
as a sum over classical paths going through the aperture at
different times depending on the position at which the particle
passes through the slit. To illustrate the semi-classical
transition we calculate numerically the probability density
function for a narrow Gaussian wave packet, $\sigma\sim0$. In
Section IV we give a correction to the truncation approximation
propagator in the Fraunhofer regime when the dimensions of the
aperture are small compared to the distance between the slit and
the screen. Then we give a formula for the shift in the distance
between two successive minima of the probability distribution
function compared to the classical result. 
In the last section, we discuss an experimental perspective to the
diffraction pattern for a relatively large aperture of the slit,
particularly for the slit diffraction experiment in the presence of a constant gravity field.

\section{Diffraction in space (DIS) and in time (DIT) of a localized wave packet}

The aim of this section is to recall the theory of diffraction in
space and time and to give a general solution to the Schr\"odinger
equation for an initial Gaussian wave packet on the half space
delimited by a plane. The purpose of this study is to give the
physical ingredients and the mathematical tools to treat the
problem of the slit diffraction beyond the truncation
approximation. The latter main problem will be explored in the
subsequent sections.

\subsection{Basic set up}

The diffraction-in-time experiment consists of opening a shutter
at position $x_1=0$ at a time $t_1 \geq 0$ and observing the
particle at a point $x>0$ after the opening time $t-t_1>0$. In
\cite{Moshinsky} as well as in \cite{Zeilinger}, the wave at the
source is considered to be a monochromatic plane wave. Here we
consider, as in \cite{Goussev}, a localized wave packet
(Gaussian), but we follow the method developed in \cite{Zeilinger}
to find the general solution. To understand the difference between
the localized wave packet versus plane wave, we notice that the
phase of the wave is non-linear in space and in time (for one
dimension $\varphi_t(x)=\frac{m x^2}{2\hbar t}$) and so the coordinate and
time of emission of the localized wave has to be taken into
account (which is not the case for a plane wave since the phase is
linear in time, $\varphi_t(x)=kx-\frac{\hbar k^2t}{2m}$). Thus, for the
truncation approximation model, the half-plane diffraction
amplitude for a Gaussian wave packet is given by Fresnel integrals
(see \cite{Beau}) whereas for a plane wave this amplitude is given
by the Fourier transform of the shape of the aperture (for example
of a two dimensional gate function for a rectangular aperture). We
will see that the result for the space diffraction of a localized
wave packet by an half-plane is actually similar to the so-called
``diffraction in time''.

To give a general solution of the diffraction in space and time
problem, we first write the Schr\"odinger equation for the wave
function of the particle moving in the apparatus:
\begin{equation}\label{SchEq}
\begin{array}{ll}
\frac{\hbar^2}{2m}\nabla^2 \psi(\bold{r},t)+i\hbar\frac{\partial}{\partial t}\psi(\bold{r},t) = 0  \\
\psi(\bold{r},t) = 0 \mbox{ for } x
> x_1 \mbox{ and } t< t_1, \mbox{ and }
\psi(\bold{r}_1,t)=\phi(\bold{r}_1,t) \mbox{ for } t > t_1.
\end{array}
\end{equation}
Here we fixed the initial condition in the half-plane to be zero
at times $t < t_1$ and inhomogeneous Dirichlet boundary conditions
on the plane of the slit $x=x_1$ for $t > t_1$.
In \cite{Zeilinger}, the boundary condition is taken to be a
monochromatic plane wave $\phi(\bold{r}_1,t)=e^{-i\omega_0t}$, 
whereas here we will consider a localized wave packet (Gaussian).

We would like to write the solution of (\ref{SchEq}) using the
point source method by computing the Green function solution of
the equation:\cite{Morse}
\begin{equation}\label{greenE}
\frac{\hbar^2}{2m}\nabla^2 G(\bold{r},t,\bold{r'},\tau)+i\hbar\frac{\partial G(\bold{r},t,\bold{r'},\tau)}{\partial t}
= i\hbar\delta^3(\bold{r}-\bold{r'})\delta(t-\tau)
\end{equation}
with the causality conditions:
\begin{equation}\label{greenC}
G(\bold{r},t<\tau,\bold{r'},\tau)=0,\ \nabla G(\bold{r},t<\tau,\bold{r'},\tau)=0
\end{equation}

The free Green function for infinite volume with the conditions (\ref{greenC}) is:
\begin{equation}\label{green0}
G_0(\bold{r}-\bold{r'};t-\tau)=\left(\frac{m}{2i\pi\hbar(t-\tau)}\right)^{3/2}e^{\frac{im|\bold{r}-\bold{r'}|^2}{2\hbar(t-\tau)}}
\theta(t-\tau)
\end{equation}

Here we recall the solution of the Schr\"odinger equation
(\ref{SchEq}) and we refer the reader to \cite{Zeilinger} and
\cite{Morse} for more details:
\begin{multline}\label{SchSol1}
\psi(\bold{r},t)= \int_V d^3\bold{r}' G(\bold{r},t,\bold{r}^{'},t_1)\psi(\bold{r}^{'},t_1)\\
+ \frac{i\hbar}{2m}\int_{t_1}^{t} d\tau \int_{\partial V}d\bold{S}_1
\left[ G(\bold{r},t,\bold{r}_1,\tau)\nabla_{\bold{r}_1}\psi(\bold{r}_1,\tau)
- \psi(\bold{r}_1,\tau) \nabla_{\bold{r}_1}G(\bold{r},t,\bold{r}_1,\tau) \right]
\end{multline}
Here $\partial V$ is the boundary of the half-plane, i.e. the
2-dimensional surface $x=x_1$.

In the following, we denote by $\bold{r}_{\perp}=(y,z)$ the
coordinates in the plane orthogonal to the $x$-axis and
$\bold{r}_{\perp,1}=(y_1,z_1)$ the same at the shutter. We
consider general homogeneous conditions for the Green function:
\begin{equation}\label{greenG}
G(\bold{r},t,\bold{r}_1,\tau)=\lambda_1 G_0(x-x_1,\bold{r}_{\perp}-\bold{r}_{\perp,1};t-\tau)
+\lambda_2 G_0(x+x_1,\bold{r}_{\perp}-\bold{r}_{\perp,1};t-\tau)\ .
\end{equation}
By a direct calculus we have :
\begin{equation}\label{greenGder}
\partial_{x_1} G(\bold{r},t,\bold{r}_1,t_1)|_{x_1=0}
= \left( -\lambda_1 + \lambda_2 \right)\frac{im}{\hbar}\frac{x}{t-\tau} G_0(x,\bold{r}_{\perp}-\bold{r}_{\perp,1};t-t_1)
\end{equation}
In particular we have the following special cases:\\
(i) for $\lambda_1=1\, \lambda_2=-1$, we have the homogeneous Dirichlet conditions, $G(\bold{r},t,\bold{r}_1,\tau)|_{x_1=0}=0$\\
(ii) for $\lambda_1=1\, \lambda_2=1$, we get the homogeneous Neumann conditions,\\ $\partial_{x_1}G(\bold{r},t,\bold{r}_1,\tau)|_{x_1=0}=0$\\
(iii) for $\lambda_1=1\, \lambda_2=0$, we get the free Green's function $G(\bold{r},t,\bold{r}_1,\tau)=G_{0}(\bold{r}-\bold{r}_1;t-\tau)$\\

Notice that the volume is the half-space to the right-hand side of
the shutter $V=[0,+\infty)\times \mathbb{R}\times\mathbb{R}$ and
we consider that the initial wave function vanishes in this domain
$\psi(\bold{r}',t_1)=0$, if $x'>0$, with $\bold{r}'=(x',y',z')$.
Then by (\ref{SchSol1}), we get the following solution :
\begin{equation}\label{PsiS1}
\psi(\bold{r},t) = \frac{i\hbar}{2m}\int_{t_1}^{t} d\tau
\int_{\partial V}d\bold{S}_1 \left[
G(\bold{r},t,\bold{r}_1,\tau)\nabla_{\bold{r}_1}\psi(\bold{r}_1,\tau)
- \psi(\bold{r}_1,\tau)
\nabla_{\bold{r}_1}G(\bold{r},t,\bold{r}_1,\tau) \right]_{x_1=0}.
\end{equation}
Note that $d\bold{S}_1$ is the elementary boundary surface vector
orthogonal to the plane at the point $\bold{r}_1$ and pointing
outward of the volume (i.e. $d\bold{S}_1=-dy_1 dz_1 \bold{e}_x$).
On the surface of the aperture $\partial V$, we consider that after opening the
shutter, the wave function is a Gaussian wave packet which was
emitted at time $t_0=0$, and therefore given by the following wave
function at each point $\bold{r}_1\in
\partial V$:
\begin{equation}\label{Psit1}
\psi(\bold{r}_1,\tau) = \int_{\mathbb{R}^3}d\bold{R}\ G_{0}(\bold{r}_1-\bold{R};\tau) \phi(\bold{R},0)\theta(\tau-t_1) \ ,
\end{equation}
where the normalized Gaussian wave packet $\phi$ is given by:
\begin{equation}\label{Psi0}
\phi(\bold{R},0)=\frac{1}{(2\pi\sigma^2)^{3/4}}e^{-\frac{|\bold{R}-\bold{r}_0|^2}{4\sigma^2}} \ ,
\end{equation}
where $\bold{R}=(X,Y,Z)$ and so $X$ denotes the coordinate along
the $x$-axis. The probability density for the initial wave packet
is such that $|\phi(\bold{R},0)|^2\rightarrow
\delta^3(\bold{R}-\bold{r'})$ when $\sigma\rightarrow0$. In the
sequel we will consider the case that $\sigma$ is small compared
to the distance $|x_1-x_0|$ between the position $x_0$ of the
center of the Gaussian of the wave packet
and the position of the shutter $x_1$.\\

\textit{Remark}. {To relate the conditions (\ref{Psit1}) to the condition in \cite{Zeilinger}, let us rewrite
the initial condition at the emission of the wave packet as a Gaussian distribution of plane waves:
\begin{equation}\label{GaussDecomp}
\phi(\bold{R},0)=\int_{\mathbb{R}^3}d\bold{k}\ \varphi_{\bold{k}}(\bold{r}_0-\bold{R},0)e^{-\frac{\sigma^2}{2}\bold{k}^2}
\end{equation}
where
$\phi_{\bold{k}}(\bold{r}_0-\bold{R},0)=e^{i\bold{k}\cdot(\bold{r}_0-\bold{R})}$.
Then, if we choose the same boundary condition on the surface
$(x_1=0,y_1,z_1)$ for the plane waves defined just above as the
one considered in \cite{Zeilinger}:
\begin{equation}\label{ZBC}
\varphi_{\bold{k}}(\bold{r}_0-\bold{R},\tau)=e^{i\bold{k}\cdot(\bold{r}_1-\bold{r}_0)}e^{-i\omega \tau}\theta(\tau-t_1)
\end{equation}
with the dispersion relation $\omega=\frac{\hbar\bold{k}^2}{2m}$, we
directly get (\ref{Psit1}) from (\ref{ZBC}) and
(\ref{GaussDecomp}). Notice that in (\ref{Psi0}) we have
arbitrarily chosen the initial wave vector to be zero, but we
could generally set:
\begin{equation}\label{Psik0}
\phi_{\bold{k}_0}(\bold{R},0)=\frac{1}{(2\pi\sigma^2)^{3/4}}e^{-\frac{|\bold{R}-\bold{r}_0|^2}{4\sigma^2}}e^{i\bold{k}_0\cdot(\bold{r}_0-\bold{R})} \ ,
\end{equation}
where the initial wave vector is $\bold{k}_0$. However, in the
following, we will assume that $\sigma$ is close to zero and so
there will not be a privileged initial wave vector, which is why
we take $\bold{k}_0=\bold{0}$ in the sequel.} \medskip

By (\ref{PsiS1}) and (\ref{Psit1}), we get the following formula
\begin{equation}\label{PsiS2}
\psi(\bold{r},t) = \int_{R^3} d\bold{R}\ K(\bold{r},t,\bold{R},0|\partial V,t_0)  \phi(\bold{R},0)
\end{equation}
where the propagator is defined by:
\begin{multline}\label{K}
K(\bold{r},t,\bold{R},0|\partial V,t_1)\equiv \\
\frac{i\hbar}{2m}\int_{t_1}^{t} d\tau \int_{\partial V}d\bold{S}_1 \cdot
\left[ G(\bold{r},t,\bold{r}_1,\tau)\nabla_{\bold{r}_1}G_0(\bold{r}_1-\bold{R},\tau)
- G_0(\bold{r}_1-\bold{R},\tau) \nabla_{\bold{r}_1}G(\bold{r},t,\bold{r}_1,\tau) \right]_{x_1=0}
\end{multline}
\textit{Remark}. To avoid confusion, we stress that (\ref{PsiS2})
is different from the volume integral term of the general solution
(\ref{SchSol1}): we have just rewritten (\ref{PsiS1}) using the
expression (\ref{K}) and the integral (\ref{Psit1}).

Since
\begin{gather*}
\nabla_{\bold{r}_1}G_0(\bold{r}_1-\bold{R},\tau)=\frac{im}{\hbar}\frac{\bold{r}_1-\bold{R}}{\tau}G_0(\bold{r}_1-\bold{R},\tau)\\
\nabla_{\bold{r}_1}G(\bold{r},t,\bold{r}_1,\tau)
=\frac{im}{\hbar}\left(-\lambda_1\frac{\bold{r}-\bold{r}_1}{t-\tau} + \lambda_2\frac{\bold{r}+\bold{r}_1}{t-\tau}\right)G_0(\bold{r}-\bold{r}_1,\tau)
\end{gather*}
we get:
\begin{multline}\label{K1}
K(\bold{r},t,\bold{R},0|\partial V,t_1)=
-\frac{1}{2}\int_{t_1}^{t} d\tau \int_{\partial V}d\bold{S}_1 \cdot\Big(\\
\left[\frac{\bold{r}_1-\bold{R}}{\tau}(\lambda_1+\lambda_2)
+\lambda_1\frac{\bold{r}-\bold{r}_1}{t-\tau} - \lambda_2\frac{\bold{r}+\bold{r}_1}{t-\tau}\right]_{x_1=0}
G_0(\bold{r}-\bold{r}_1,t-\tau) G_0(\bold{r}_1-\bold{R},\tau)|_{x_1=0}\Big)
\end{multline}

\subsection{One-dimensional diffraction in time of a localized wave packet}

Consider the one-dimensional diffraction-in-time problem for a
Gaussian wave packet emitted at $x_0<0$ at the time $0$. By
similar arguments to those leading to (\ref{K1}), we get, for
general boundary conditions, the following propagator:
\begin{equation}\label{K(G)}
K(x,t,X,0|x_1=0,t_1) = \int_{t_1}^t d\tau
\left[\frac{-X}{\tau}\eta_{1}+\frac{x}{t-\tau}\eta_{2}\right]G_0(x,t-\tau)G_0(-X,\tau)
\end{equation}
where we put
\begin{eqnarray}\label{eta}
&& \eta_{1}=\frac{1}{2}\left(\lambda_1+\lambda_2\right) {}\\{}&&
\eta_{2}=\frac{1}{2}\left(\lambda_1-\lambda_2\right).
\end{eqnarray}
Notice that choosing $\eta\equiv\eta_{2}=1-\eta_{1}\
,\eta\in\mathbb{C}$ and taking $t_1=0$, a direct calculation shows
that the integral in (\ref{K(G)}) is equal to $G_{0}(x-X;t)$ and
so it gives the general solution for the free particle motion, see
\cite{Goussev}. Now, if we take $\eta=1/2$ (i.e. $\lambda_2=0$ and
$\lambda_1=1$) then we get the free boundary condition corresponding to
the perfectly absorbing shutter-screen condition. The correct solution 
for $t_1>0$ is equivalent to the Moshinsky solution:
\begin{equation}\label{K(0)1}
K^{(0)}(x,t,X,0|x_1=0,t_1) = \frac{1}{2}\int_{t_1}^t d\tau\
\left[-\frac{X}{\tau}+\frac{x}{t-\tau}\right]G_0(x,t-\tau)G_0(-X,\tau)
\end{equation}
which is easily evaluated (see \cite{Goussev}):

\begin{multline}\label{K(0)2}
K^{(0)}(x,t,X,0|x_1=0,t_1)=\\
G_{0}(x-X;t)
\left(1+
\frac{1}{2}\mathrm{erfc}
\left(
\left(x\frac{t_1}{t}+X\frac{t-t_1}{t}\right)
\left(\frac{m t}{2i\hbar t_1(t-t_1)}\right)^{1/2}
\right)
\right)
\end{multline}

Hence, by (\ref{K(G)}), we get the propagator for general
homogeneous boundary conditions,
\begin{equation}
K^{(G)}(x,t,X,0|x_1=0,t_1)=\lambda_1 K^{(0)}(x,t,X,0|x_1=0,t_1)
-\lambda_2 K^{(0)}(x,t,-X,0;x_1=0,t_1)\ .
\end{equation}
similar to the case of a monochromatic plane wave
\cite{Zeilinger}. For $\lambda_1=-\lambda_2=+1$ we get Dirichlet
boundary conditions whereas for $\lambda_1=\lambda_2=+1$ we have
Neumann boundary conditions.

The solution of the one-dimensional Schr\"odinger equation for the perfectly absorbing shutter-screen is
obtained by inserting (\ref{K(0)1}) into the one-dimensional version of (\ref{PsiS2}):
\begin{multline}\label{psi(x,t)}
\psi(x,t)=\int_{-\infty}^{+\infty}dX\ K^{(0)}(x,t;X,0|x_1=0,t_1)\psi(X,0)\\
=\int_{-\infty}^{+\infty}dX\
K^{(0)}(x,t;X,0|x_1=0,t_1)\frac{e^{-\frac{(X-x_0)^2}{4\sigma^2}}}{(2\pi\sigma^2)^{1/4}}.
\end{multline}
In particular, if we assume that $\sigma\ll |x_0|$, we get that
$\psi(x,t)\approx (8\pi\sigma)^{1/4}K^{(0)}(x,t;x_0,0|x_1=0,t_1)$.

Notice that the explicit solution (\ref{K(0)2}) for $\sigma\ll
|x_0|$ is similar to the explicit formulas giving the propagator
and the wave function for the half-plane diffraction problem in
the truncation approximation \cite{Goussev}. So both diffraction phenomena
are analogous and this is why we use the term ``diffraction in
time'' even for the localized wave packet. In the next subsection,
we will see that we can also construct an analogous Moshinsky
shutter problem for a localized wave packet and show the
equivalence between both approaches for general homogeneous
boundary conditions.

\section{One-slit diffraction model and its semi-classical approximation}

In the last section, we gave the theory of diffraction in space
and time for a wave packet and we furnished the general solution
of the Schr\"odinger equation for an initial Gaussian wave packet
passing through an aperture which is opened at a time $t_1\geq
t_0=0$, where $t_0$ is the time of the emission of the initial
wave packet. We have also seen that we can interpret this
phenomenon as a diffraction in time by analogy with the
diffraction in space. However, since the main problem of this
article is to derive a formula for the slit diffraction problem,
where the aperture is not assumed to be small compared with the
distance between the slit and the screen (and the slit and the
source), we would like to interpret the so-called diffraction in
time phenomenon in a different way, where the apparatus is fixed
in time (no shutter) and the problem is stationary. In this
section, we apply the theory developed in the previous
section to the slit diffraction problem and give a geometric
interpretation for the propagator. We first give an explicit
formula for the propagator with general boundary conditions, then
give its semi-classical expression and comment on the results. We
also give numerical results for intensity patterns on the
screen in the delta limit $\sigma\rightarrow0$ for the Dirichlet,
Neumann and free boundary conditions and comment on the
differences. In the next section, we will use the semi-classical
formula of the propagator to give corrections to the truncation
approximation model.

\subsection{Single-slit diffraction of a narrow Gaussian wave packet}

We consider the slit $\Omega_{a,b}\equiv\{x_1=0\}\times
[-b,b]\times [-a,a]$, and assume that the shutter is open at the
time $t_1=0$. The dynamics of the particle obeys the Schr\"odinger
equation (\ref{SchEq}) and the boundary conditions are given by
(\ref{Psit1}), (\ref{Psi0}). Here we consider the general
homogeneous boundary conditions (\ref{greenG}). By (\ref{K1}) we
get the following formula for the propagator:
\begin{equation}\label{KSlit1}
K^{(a,b)}(\bold{r},t,\bold{R},0) = \int_0^t d\tau \int_{-a}^{a}dz_1\int_{-b}^{b}dy_1
\left[\frac{-X}{\tau}\eta_{1}+\frac{x}{t-\tau}\eta_{2}\right]G_0(\bold{r}-\bold{r}_1,t-\tau)G_0(\bold{r}_1-\bold{R},\tau)
\end{equation}

The integral over the time of the one-point source propagator (for
every $\bold{r}_1\in \partial V$ fixed) can be evaluated
explicitly. The resulting formula will be analyzed in the
semi-classical limit using  the stationary phase approximation
method which yields a semi-classical interpretation of the
propagator. The one-slit propagator is then given by  an integral
of this one-point source propagator over the aperture of the slit.
Finally, the wave function on the screen at time $t$ is given by
(\ref{PsiS2}), where we consider an initial narrow Gaussian wave
packet (\ref{Psi0}) of width $\sigma$ which is small compared to
the distance between the center of the Gaussian and the slit and
also to the size of the aperture:
\begin{gather*}
\sigma\ll |\bold{r}_0|,\ a,b.
\end{gather*}
By (\ref{PsiS2}) we then have the following approximation:
\begin{gather}
\psi(\bold{r},t) = (8\pi\sigma^2)^{3/4}\int_{R^3} d\bold{R}\ K^{(a,b)}(\bold{r},t,\bold{R},0)
\frac{e^{-\frac{|\bold{R}-\bold{r}_0|^2}{4\sigma^2}}}{(2\pi\sigma^2)^{3/2}}\nonumber \\
\approx(8\pi\sigma^2)^{3/4}K^{(a,b)}(\bold{r},t,\bold{r}_0,0),\ \textrm{when}\ \sigma\sim0\label{PsiS3}
\end{gather}
Therefore, in the sequel we will take $\bold{R}=\bold{r}_0$ in (\ref{KSlit1})
since the final wave function is just proportional to the one-slit propagator.\\

\textit{Remark}. In the limit $\sigma\rightarrow 0$, we would like
to give a formula for the probability density for the particle to
be at the point $\bold{r}$ on the screen at the time $t$. This has
already been done for the truncation approximation, see
\cite{Beau} and Appendix 1, and the general idea here is similar. It
is important to realize that $|\psi(\bold{r},t)|^2$ represents the
non-normalized wave function at the point $\bold{r}$ on the screen
at the time $t$ and so, to get the probability, we have to divide
by the total mass on the screen:
$$ M\equiv\int_{-\infty}^{+\infty}dy\int_{-\infty}^{+\infty}dz\ |\psi(\bold{r},t)|^2\ .$$
For the truncation approximation model, the particle is assumed to
pass through the slit at the classical time $t_c$ (given by a
linear relation between $t$ and the distances along the $x$-axis).
Hence, the total mass passing to the right side of the plane of
the slit is equal to the total mass on the screen:
\begin{equation}\label{PCLaw}
\int_{- b}^{b}dy_1\int_{-a}^{a}dz_1\ |\psi_{\textrm{Trunc}}(\bold{r}_1,t_c)|^2
=\int_{-\infty}^{+\infty}dy\int_{-\infty}^{+\infty}dz\ |\psi_{\textrm{Trunc}}(\bold{r},t)|^2
\end{equation}
However, in our model there is no similar conservation equation to
(\ref{PCLaw}) since we do not know the exact time when the
particle passes through the aperture. Hence, the expression for
the probability density  at the point $\bold{r}$ and at the time
$t$ has to written as
\begin{equation}\label{Plim}
P(\bold{r},t)= \frac{1}{M}|\psi(\bold{r},t)|^2\rightarrow\frac{1}{\Omega}
|K^{(a,b)}(\bold{r},t,\bold{r}_0,0)|^2,\ \textrm{when}\ \sigma\rightarrow0
\end{equation}
where $\Omega\equiv
\int_{-\infty}^{+\infty}dy\int_{-\infty}^{+\infty}dz|K^{(a,b)}(\bold{r},t,\bold{r}_0,0)|^2$.\\

The formula (\ref{KSlit1}) can be rewritten as an integral over
all points $\bold{r}_1=(x_1,y_1,z_1)$ in the slit (i.e.
$y_1\in[-b,b]$ and $z_1\in[-a,a]$):
\begin{equation}\label{KRew}
K^{(a,b)}(\bold{r},t,\bold{r}_0,0)
=\int_{-a}^{a}dz_1\int_{-b}^{b}dy_1 K(\bold{r},t;\bold{r}_0,0|\bold{r}_1)\ ,
\end{equation}
where we have defined the three-dimensional one-point source propagator:
\begin{equation}\label{KS1}
K(\bold{r},t;\bold{r}_0,0|\bold{r}_1)\equiv \int_0^t d\tau
\left[\frac{-x_0}{\tau}\eta_{1}+\frac{x}{t-\tau}\eta_{2}\right]G_0(\bold{r}-\bold{r}_1,t-\tau)G_0(\bold{r}_1-\bold{r}_0,\tau)\ .
\end{equation}
We want to give an explicit formula for the one-point slit
propagator (\ref{KS1}). For a detailed calculuation we refer the
reader to Appendix 2. The result is the following explicit formula:
\begin{equation}\label{KSsol}
K(\bold{r},t;\bold{r}_0,0|\bold{r}_1)=
A_t(\bold{r};\bold{r}_0|\bold{r}_1)
e^{i\varphi_t(\bold{r};\bold{r}_0|\bold{r}_1)}
\end{equation}
where the phase is given by
\begin{equation}\label{Phase}
\varphi_t(\bold{r},\bold{r}_0|\bold{r}_1) \equiv\frac{m}{2\hbar
t}(|\bold{r}-\bold{r}_1|+|\bold{r}_1-\bold{r}_0|)^2
\end{equation}
and where the amplitude is given by a linear combination of the
Neumann and Dirichlet amplitudes:
\begin{equation}\label{A}
A_t(\bold{r},\bold{r}_1-\bold{r}_0)\equiv
\eta_{1}A_t^{(N)}(\bold{r},\bold{r}_1-\bold{r}_0)
+\eta_{2}A_t^{(D)}(\bold{r},\bold{r}_1-\bold{r}_0).
\end{equation}
The Neumann part is given by:
\begin{equation}\label{AN}
A_t^{(N)}(\bold{r},\bold{r}_0|\bold{r}_1)
= \frac{-x_0}{(2i\pi\hbar t/m)^{3/2}}\left(\frac{m}{2i\pi\hbar t}\frac{(|\bold{r}-\bold{r}_1|+|\bold{r}_1-\bold{r}_0|)^2}
{|\bold{r}-\bold{r}_1||\bold{r}_1-\bold{r}_0|^2}
+\frac{1}{2\pi|\bold{r}_1-\bold{r}_0|^3}\right)
\end{equation}
and the Dirichlet part by:
\begin{equation}\label{AD}
A_t^{(D)}(\bold{r},\bold{r}_0|\bold{r}_1) = \frac{x}{(2i\pi\hbar
t/m)^{3/2}}\left(\frac{m}{2i\pi\hbar
t}\frac{(|\bold{r}-\bold{r}_1|+|\bold{r}_1-\bold{r}_0|)^2}
{|\bold{r}-\bold{r}_1|^2|\bold{r}_1-\bold{r}_0|}
+\frac{1}{2\pi|\bold{r}-\bold{r}_1|^3}\right).
\end{equation}

\textit{Remark.} The equation (\ref{KRew}) gives the correct
propagator formula for the slit diffraction problem, whatever the
initial condition for the wave at $t_0=0$. For example, we can
take the following more general condition than (\ref{Psik0}):
$$ \phi_{\bold{k}_0}(\bold{R}=(X,Y,Z),0)=\frac{1}{((2\pi)^3\sigma_x^2\sigma_y^2\sigma_z^2)^{1/4}}
e^{-\frac{|X-x_0|^2}{4\sigma_x^2}}e^{-\frac{|Y-y_0|^2}{4\sigma_y^2}}e^{-\frac{|Z-z_0|^2}{4\sigma_z^2}}
e^{i\bold{k}_0\cdot(\bold{r}_0-\bold{R})} \ , $$ where
$\sigma_y,\ \sigma_z$ are small but $\sigma_x$ is large, and
consider in the limit delta-distributions along the $y$- and
$z$-axis and a plane wave along the $x$-axis. Then, the
approximation we made for $\sigma\sim0$ is still valid in the
plane of the slit but not on the $x$-axis which can be considered
the \textit{propagation axis}. In this limit, to get the wave
solution, we have to compute the Fourier transform of
(\ref{KRew}) with respect to $x_0$:
$$\psi(\bold{r},t)\approx \left(32 \pi \frac{\sigma_y^2 \sigma_z^2}{\sigma_x^2} \right)^{1/4}
\int_{-\infty}^{+\infty}dx_0\ K^{(a,b)}(\bold{r},t,\bold{R},0)\ e^{ik_{0,x} x_0} $$

\subsection{Semiclassical limit of the one-slit propagator}

In the following we still assume that the opening of the aperture
of the slit coincides with the emission of the Gaussian wave
packet $t_1=t_0=0$, and that the width of the Gaussian is small
compared to the distances of the apparatus, $\sigma\ll|x_0|,a,b$,
so that (\ref{PsiS3}) gives a good approximation to the solution.

Now we will give the semi-classical approximation of the propagator
(\ref{KRew}) when the fluctuation of the phase tends to zero, i.e.
considering that $\mu\equiv \frac{m|\bold{r}|^2}{\hbar t}\gg1$ and
so that $|\bold{r}|\gg \lambda_0$ with
$\lambda_0\equiv\sqrt{2\pi\hbar t/m}$. This allows us to interpret
the propagator of the slit experiment in this regime as the sum
over classical paths starting from $\bold{r}_0$ at the time
$t_0=0$ to $\bold{r}$ at the time $t$ given that the particle
passes through the slit $\bold{r}_1\in \Omega_{a,b}$ at a
so-called \textit{semi-classical time} $\tau_{sc}\in(0,t)$. The
results are similar to the ones obtained for the truncation
approximation model (see \cite{Beau} and Appendix 1) but not the
same since we do not make any geometrical approximation and, as a
consequence, $\tau_{sc}$ depends on the coordinates $\bold{r}_1$
of the classical paths passing through the slit. We should mention
that another condition for the validity of the semi-classical
approximation is that the distances $|\bold{r}-\bold{r}_1|$ and
$|\bold{r}_1-\bold{r}_0|$ have to be of the same order for every
$(y_1,z_1)\in\Omega_{a,b}$, which means that $|x_0|$ and $|x|$ are
also of the same order, and that the sizes of the aperture $a,\ b$
and the position of the screen $|y|,\ |z|$
have to be at most of the same order as $|x|$. \\

By the above assumptions, we are able to use the stationary phase
approximation applied to the one-point propagator formula(\ref{KS1}):
\begin{equation}\label{SP0}
\int_0^t d\tau f(\tau) e^{i\mu \phi(\tau)}
\approx f(\tau_{sc})e^{i\mu\phi(\tau_{sc})}\int_0^t d\tau\ e^{\frac{i\mu}{2}\phi^{''}(\tau_{sc})(\tau-\tau_{sc})^2},\ \mu\gg1
\end{equation}
where $\tau_{sc}$ is the solution of the equation
$\phi^{'}(\tau)=0$, $\phi^{''}(\tau_{sc})$ is the second
derivative of $\phi$ at the point $\tau_{sc}$, and where we put
\begin{equation}\label{SP1}
\begin{array}{ll}
f(\tau)=\frac{1}{((2i\pi\hbar/m)^2(t-\tau)\tau)^{3/2}}\left(\frac{-x_0}{\tau}\eta_{1}+\frac{x}{(t-\tau)}\eta_{2}\right)\\
\mu=\frac{m|\bold{r}|^2}{2\hbar t}=\frac{\pi |\bold{r}|^2}{\lambda_0^2}\\
\phi(\tau)=\frac{|\bold{r}-\bold{r}_1|^2}{|\bold{r}|^2(1-\tau/t)}
+\frac{|\bold{r}_1-\bold{r}_0|^2}{|\bold{r}|^2\tau/t}
\end{array}
\end{equation}

By a direct calculation, we find that the saddle point $\tau_{sc}$
which is the solution of the equation $\phi^{'}(\tau)=0$ is given by
\begin{equation}\label{tc}
\tau_{sc}=\frac{|\bold{r}_1-\bold{r}_0|}{|\bold{r}-\bold{r}_1|+|\bold{r}_1-\bold{r}_0|}t
\end{equation}
Then we get:
\begin{equation}\label{SP2}
\begin{array}{ll}
f(\tau_{sc})=\frac{1}{((2i\pi\hbar/m)^2(t-\tau_{sc})\tau_{sc})^{3/2}}\left(\frac{-x_0}{\tau_{sc}}\eta_1+\frac{x}{(t-\tau_{sc})}\eta_2\right)\\
\mu\phi(\tau_{sc})=\frac{m|\bold{r}-\bold{r}_1|^2}{2\hbar(t-\tau_{sc})}+\frac{m|\bold{r}_1-\bold{r}_0|^2}{2\hbar\tau_{sc}}
=\frac{m}{2\hbar t}\left(|\bold{r}-\bold{r}_1|+|\bold{r}_1-\bold{r}_0|\right)^2\\
\mu\phi^{''}(\tau_{sc})=\frac{m}{\hbar}\left(\frac{|\bold{r}-\bold{r}_1|^2}{(t-\tau_{sc})^3}
+\frac{|\bold{r}_1-\bold{r}_0|^2}{\tau_{sc}^3}\right)
=\frac{m}{\hbar}\frac{t^3}{(t-\tau_{sc})^3\tau_{sc}^3}\frac{|\bold{r}-\bold{r}_1|^2|\bold{r}_1-\bold{r}_0|^2}
{(|\bold{r}-\bold{r}_1|+|\bold{r}_1-\bold{r}_0|)^2}
\end{array}
\end{equation}
To estimate the integral at the right hand side of (\ref{SP0}), we
need to integrate in the complex plane along a contour, which we
take to be the perimeter of the eighth part of a circle centered
at $-\tau_{sc}$ on the real axis and of radius $t$, together with
the radii. Putting $ N\equiv \mu\phi^{''}(\tau_{sc})$, we get the
following estimate for large $N$:
\begin{multline}\label{Estimate}
\int_{-\tau_{sc}}^{t-\tau_{sc}}ds\ e^{i\frac{N s^2}{2}}=e^{i\frac{\pi}{4}}\int_{-\tau_{sc}}^{t-\tau_{sc}}ds\ e^{-\frac{N s^2}{2}}
+ it\int_{0}^{\frac{\pi}{4}}d\theta\ e^{i\theta}e^{-\frac{N t^2}{2} e^{2i\theta}}
=(\frac{2i\pi}{N})^{1/2}+O(\frac{1}{N})\ ,
\end{multline}
since the latter integral is of the order $1/N$.

Hence by (\ref{SP0}) and (\ref{Estimate}), we get the following approximation for the one-point propagator:
\begin{multline}
K_t(\bold{r},\bold{r}_0|\bold{r}_1)
\approx f(\tau_{sc})\left(\frac{2i\pi}{\mu\phi^{''}(\tau_{sc})}\right)^{1/2}e^{i\mu\phi(\tau_{sc})}\\
=\frac{(|\bold{r}-\bold{r}_1|+|\bold{r}_1-\bold{r}_0|)^2}
{(2i\pi \hbar t/m)^{5/2}|\bold{r}-\bold{r}_1|\times|\bold{r}_1-\bold{r}_0|}
\left(\frac{-x_0}{|\bold{r}_1-\bold{r}_0|}\eta_{1} + \frac{x}{|\bold{r}-\bold{r}_1|}\eta_{2}\right)
e^{\frac{im}{2\hbar t}\left(|\bold{r}-\bold{r}_1|+|\bold{r}_1-\bold{r}_0|\right)^2}\ .\label{SC1}
\end{multline}
The stationary phase approximation method, leading to the formula
(\ref{SC1}), thus yields the propagator (\ref{KSsol}) except for
the last terms of (\ref{AD}) and (\ref{AN}). Indeed, the phase is
the same and the amplitude is a linear combination of the first
term of the amplitudes (\ref{AD}), (\ref{AN}). We can explain this
result remarking that both first terms in (\ref{AD}) and
(\ref{AN}) are large compared to the second terms since the ratio
is of the order $m|\bold{r}|^2/\hbar t \gg 1$.

Rewriting the semi-classical propagator (\ref{SC1}) using
(\ref{SP2}) we have
\begin{equation}\label{SC2}
K_t^{(sc)}(\bold{r},\bold{r}_0|\bold{r}_1)
=\sigma_{t,\tau_{sc}}(x,x_0)\frac{e^{\frac{im x^2}{2\hbar
t}}}{(2i\pi\hbar t/m)^{1/2}} \frac{e^{\frac{im
[(y-y_1)^2+(z-z_1)^2]}{2\hbar(t-\tau_{sc})}}}{2i\pi\hbar
(t-\tau_{sc})/m} \frac{e^{\frac{im
[y_1^2+z_1^2]}{2\hbar\tau_{sc}}}}{2i\pi\hbar\tau_{sc}/m},
\end{equation}
where the function $\sigma_{t,\tau_{sc}}$ is defined by
\begin{equation}\label{sigma}
\sigma_{t,\tau_{sc}}(x,x_0)\equiv\frac{\lambda_0^2}{\rho}
\left(\frac{-m x_0}{2\pi\hbar\tau_{sc}}\eta_{1}+\frac{m x}{2\pi\hbar(t-\tau_{sc})}\eta_{2}\right)
\end{equation}
and where
$\rho\equiv|\bold{r}-\bold{r}_1|+|\bold{r}_1-\bold{r}_0|$ can be
interpreted as the semi-classical path length traveled by the
particle: see Fig. 1. Hence the one-slit propagator formula
(\ref{KRew}) can be written as follows:
\begin{equation}\label{SC3}
K_{sc}^{(a,b)}(\bold{r},t;\bold{r}_0,0)
=\frac{e^{\frac{im x^2}{2\hbar t}}}{(2i\pi\hbar t/m)^{1/2}}\int_{-a}^{a}dz_1\int_{-b}^{b}dy_1\ \sigma_{t,\tau_{sc}}(x,x_0)
\frac{e^{\frac{im [(y-y_1)^2+(z-z_1)^2]}{2\hbar(t-\tau_{sc})}}}{2i\pi\hbar (t-\tau_{sc})/m}
\frac{e^{\frac{im [y_1^2+z_1^2]}{2\hbar\tau_{sc}}}}{2i\pi\hbar\tau_{sc}/m}
\end{equation}

The formula (\ref{SC3}) is similar to the one-point source
propagator in the truncation approximation, formula (\ref{Ktr}) of
Appendix 1, except that the semi-classical time $\tau_{sc}$ depends
on the distance from the origin to the point in the slit, and from
the slit to the screen (see (\ref{tc})),
and the function $\sigma_{t,\tau_{sc}}$ in front of the product of the two Gaussians depends on the boundary conditions. \\

\begin{figure}[h]
\begin{center}
\scalebox{0.5}{\includegraphics[angle=-90]{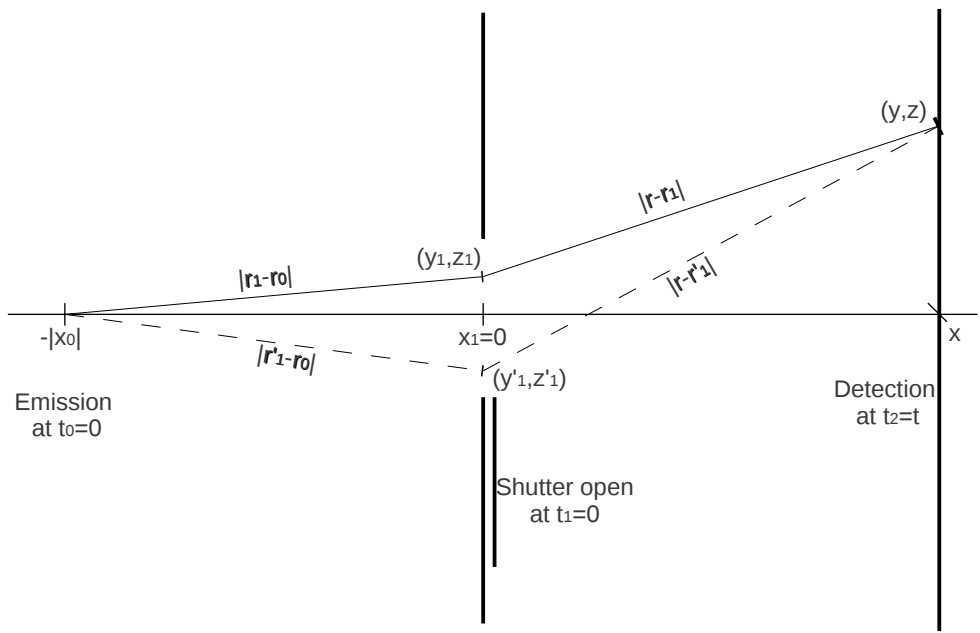}}
\caption{Schematic representation of the apparatus. We illustrate
two interfering classical paths starting from $\bold{r}_0$ at the
time $t_0=0$ to $\bold{r}$ at the time $t$ and passing through the
aperture at $\bold{r}_1$ (resp. $\bold{r}'_1$) at the time
$\tau_{sc}$ (resp. $\tau'_{sc}$) given by the formula (\ref{tc}).}
\end{center}
\end{figure}

\textit{Remark 1}. We can give a geometric interpretation to the
diffraction in space and in time in the semi-classical regime. The
first-order term of the semi-classical approximation (\ref{SC2})
gives only the classical path contribution. Therefore, we observe
that the propagator (\ref{SC3}) is nothing but a sum over all
\textit{semi-classical paths} (made up of two broken lines) 
passing through the aperture of the slit as in Fig.1. 
Thus, the equation (\ref{SC3}) shows that the
semi-classical approximation is in fact a truncation approximation
since one sees in these formulas that the motion along the
$x$-axis and the motion in the orthogonal $(y,z)$ plane are
separated and moreover that along the $x$-axis the motion is
classical. 
However, we have more information within our model
since by (\ref{tc}) we notice that there is a relation between
the classical times $\tau_{sc}$ and $t$ even if the 
two motions from the source to the slit and from the slit to the screen are separated, see Fig. 1.
Actually, we can interpret this relation as the conservation of the classical energy of the particle
when the particle passes through the slit:
$$E(\bold{r}_1,\tau_{sc})=E(\bold{r},t) \Leftrightarrow  
\frac{m}{2}\left|\frac{\bold{r}_1-\bold{r}_0}{\tau_{sc}-t_0}\right|^2
=\frac{m}{2}\left|\frac{\bold{r}-\bold{r}_1}{t-\tau_{sc}}\right|^2$$
and this leads to (\ref{tc}),
whereas the classical momentum is not conserved due to the quantum diffraction phenomenon.

Consequently, in the semi-classical regime, the theory
of diffraction in time allows us to take into account all the
classical paths (passing through the slit at a time depending on
the position inside the aperture) without assuming that the
dimensions of the aperture are small compared to the dimension of
the apparatus along the $x$-axis. Moreover, the additional term
(\ref{sigma}) which could not be discovered otherwise, has an
important physical meaning since the values of the parameters
$\eta_1,\ \eta_2$ depend on whether the screen of the slit is
reflective, absorbing or neither.

In Fig.2., Appendix 3 we show the transition 
between the quantum and the semi-classical regimes from the left to the right.  

Firstly, for the diffraction patterns at the left side (Fig.2.1a-2.4a), 
the \textit{semi-classical parameter} $\mu\sim4$ (i.e., relatively close to one)
and so that explain why the curves are different from those for the truncation approximation (see Fig.2.4a).
We observe that for the Dirichlet boundary condition (Fig.2.1a) there is a narrow central peak decreasing very fast 
so that we can not see the oscillations (a numerical zoom could show these slight oscillations).
On the contrary, for the Neumann (Fig.2.2a) and the free boundary conditions (Fig.2.3a), there is no central peak
but large oscillations where the distance between the fringes is essentially constant but different 
from the distance between the fringes in the truncation approximation (Fig.2.4a).

For the curves at right side of (Fig.2.1c-2.4c), we have $\mu\sim800\gg1$ and then we get similar pictures 
to those in the truncation approximation (Fig. 2.4c)
although there is still a difference in the location of the fringes. 
In the following section, we will give a qualitative description of those differences.
From (\ref{delta}) and (\ref{q}) we can conclude that if the first minima 
of the curves are not too different, however we observed that the second and the third differ by $50\%$.

The patterns in the middle of Fig.2, 
show the transition between the quantum and the semi-classical regimes
where a central peak appears also for the Neumann (Fig.2.5) and the free boundary conditions (Fig.2.8).\\

\textit{Remark 2}. Let us make another remark about the
probabilistic interpretation of the slit diffraction experiment in
the semi-classical limit. A consequence of the last comment about
the relation between the semi-classical approximation and the
truncation approximation is that there is an analogous equation to
(\ref{PCLaw}) giving the relation of the conservation of the
probability between the aperture of the slit and the screen:
\begin{equation}\label{ScPLaw}
\int_{- b}^{b}dy_1\int_{-a}^{a}dz_1\ |\psi_{\textrm{sc}}(\bold{r}_1,\tau_{sc})|^2
=\int_{-\infty}^{+\infty}dy\int_{-\infty}^{+\infty}dz\ |\psi_{\textrm{sc}}(\bold{r},t)|^2\equiv M_{sc}
\end{equation}
where $\tau_{sc}$ is given by (\ref{tc}) and depends on $\bold{r},\bold{r}_1,\bold{r}_0,t$,
and where the semi-classical wave functions in (\ref{ScPLaw}) are given by:
\begin{gather}\label{ScFct}
\psi_{\textrm{sc}}(\bold{r}_1,\tau_{sc})=\int_{\mathbb{R}^3}d\bold{R}\ G_{0}(\bold{r}_1-\bold{R},\tau_{sc})
\frac{e^{i\frac{|\bold{R}-\bold{r}_0|^2}{2\sigma^2}}}{(2\pi\sigma^2)^{3/2}}\\
\psi_{\textrm{sc}}(\bold{r},t)=\int_{\mathbb{R}^3}d\bold{R}\ K_{sc}^{(a,b)}(\bold{r},t,\bold{R},0)
\frac{e^{i\frac{|\bold{R}-\bold{r}_0|^2}{2\sigma^2}}}{(2\pi\sigma^2)^{3/2}}
\end{gather}
So we get the following formula for the semi-classical density of probability:
\begin{equation}\label{Psclim}
P_{sc}(\bold{r},t)= \frac{1}{M_{sc}}|\psi_{sc}(\bold{r},t)|^2\rightarrow\frac{1}{\Omega_{sc}}|K_{sc}^{(a,b)}(\bold{r},t,\bold{r}_0,0)|^2,
\ \textrm{when}\ \sigma\rightarrow0
\end{equation}
where $M_{sc}$ is defined in (\ref{ScPLaw}) and $\Omega_{sc}=\int_{- b}^{b}dy_1\int_{-a}^{a}dz_1\ |G_{0}(\bold{r}_1-\bold{R},\tau_{sc})|^2$.\\

\section{Semi-classical approximations for the slit experiment}

The equation (\ref{KRew}) gives the three-dimensional
one-gate-slit propagator as a double integral of the
three-dimensional one-point-slit propagator given by the equations
(\ref{KSsol}), (\ref{A}) and (\ref{Phase}). Despite the fact that
there is no explicit formula giving the result for the gate-slit
propagator, we can give an approximation when the size of the slit
and the distance on the screen are relatively small, in which case
it is also of interest to give an estimate for the relative shift
between the minima in the interference pattern for the Fraunhofer
regime compared with the truncation approximation.

We first want to give the semi-classical approximation of the
one-point source propagator (\ref{KSsol}) when the sizes in the
$x$-direction are relatively large compared to the sizes of the
slit and of the distances of the observation point on the screen:
$$|x-x_1|,|x_1-x_0|\gg a,b,|z|,|y|$$
and also large compared to $\lambda_0=\sqrt{2\pi\hbar t/m}$
(relatively short time). Therefore the semi-classical limit
(\ref{SC3}) is a good approximation.

By (\ref{Phase}), the phase of the propagator (\ref{KSsol}) is
given by
\begin{equation}\label{FraAp0}
\frac{\hbar}{m}\varphi_t =\frac{|\bold{r}-\bold{r}_1|^2}{2t}+
\frac{|\bold{r}_1-\bold{r}_0|^2}{2t}+\frac{|\bold{r}-\bold{r}_1||\bold{r}_1-\bold{r}_0|}{t}
\end{equation}
We denote the two-dimensional vectors of the position on the screen
$\bold{r}_{\perp}=(y,z)$, on the slit $\bold{r}_{\perp,1}=(y_1,z_1)$.
In the sequel, we take $x_1=0$, so $|x-x_1|=|x|$, $|x_1-x_0|=|x_0|$ and $y_0=0,\ z_0=0$, $x_0<0$.
The first two terms of the r.h.s. of (\ref{FraAp0}) are rewritten as
\begin{equation}\label{FraAp1}
\frac{|\bold{r}-\bold{r}_1|^2}{2t}+
\frac{|\bold{r}_1|^2}{2t} =\frac{x^2}{2t}+\frac{(\bold{r}_{\perp}-\bold{r}_{\perp,1})^2}{2t}+
\frac{x_0^2}{2t}+\frac{\bold{r}_{\perp,1}^2}{2t}
\end{equation}
Expanding to fourth order the third term of the r.h.s. of
(\ref{FraAp0}), we have
\begin{gather}
\frac{|\bold{r}-\bold{r}_1||\bold{r}_1-\bold{r}_0|}{t}=
\frac{|x||x_0|}{t}\left(1+\frac{(\bold{r}_{\perp}-\bold{r}_{\perp,1})^2}{x^2}\right)^{1/2}
\left(1+\frac{\bold{r}_{\perp,1}^2}{x_0^2}\right)^{1/2}\\
\approx\frac{|x||x_0|}{t}\left(1+\frac{(\bold{r}_{\perp}-\bold{r}_{\perp,1})^2}{2x^2}
-\frac{(\bold{r}_{\perp}-\bold{r}_{\perp,1})^4}{8x^4}\right)
\left(1+\frac{\bold{r}_{\perp,1}^2}{2x_0^2}-\frac{\bold{r}_{\perp,1}^4}{8x_0^4}\right)\\
\approx\frac{|x||x_0|}{t}\left(1+\frac{\bold{r}_{\perp,1}^2}{2x_0^2}
+\frac{(\bold{r}_{\perp}-\bold{r}_{\perp,1})^2}{2x^2}\right)
-\frac{|x||x_0|}{8t}\left(\frac{(\bold{r}_{\perp}-\bold{r}_{\perp,1})^2}{x^2}-
\frac{\bold{r}_{\perp,1}^2}{x_0^2}\right)^2\label{FraAp2}
\end{gather}
Due to (\ref{FraAp1}) and (\ref{FraAp2}) we get:
\begin{equation}\label{FraAp3}
\frac{\hbar}{m}\varphi_t \approx \frac{(x-x_0)^2}{2t}+\frac{(\bold{r}_{\perp}-\bold{r}_{\perp,1})^2}{2(t-t_c)}
+\frac{\bold{r}_{\perp,1}^2}{2t_c}
-\frac{|x||x_0|}{8t}\left(\frac{(\bold{r}_{\perp}-\bold{r}_{\perp,1})^2}{x^2}
-\frac{\bold{r}_{\perp,1}^2}{x_0^4}\right)^2
\end{equation}
where:
\begin{equation}\label{tcTrc}
t_c=\frac{|x_0|}{|x|+|x_0|}t=\frac{|x_0|}{|x-x_0|}t\ .
\end{equation}

\subsection{The truncation approximation}

Notice that due to (\ref{tc}), $t_c$ could be interpreted as the semi-classical time
at the first order approximation in the regime $|x|,|x_0|\gg |z|,|z_1|$:
\begin{equation}\label{tcFra}
\tau_{sc}=\frac{|\bold{r}_1-\bold{r}_0|t}{|\bold{r}-\bold{r}_1|+|\bold{r}_1-\bold{r}_0|}
\approx t_c,\ \mathrm{when}\ |x|,|x_0|\gg |z|,|z_1|\ .
\end{equation}

Inserting this into the amplitude of (\ref{SC2}), i.e. neglecting
the influence of the position on the screen and in the slit, we
get the following approximation:
\begin{equation}\label{AFraAp}
A_t(\bold{r},\bold{r}_0|\bold{r}_1)\approx
\sigma_{t,t_c}(x,x_0)\frac{1}{\sqrt{2i\pi\hbar t/m}} \frac{1}{(2i\pi\hbar/m)^2 (t-t_c)t_c}\ .
\end{equation}
where $\sigma_{t,t_c}(x,x_0)$ is given by (\ref{sigma}). In
(\ref{FraAp3}), we neglect the terms of  order
$O(\bold{r}_{\perp}^4)$ and $O(\bold{r}_{\perp,1}^4)$, to get
\begin{equation}\label{Fra2dAp}
\frac{\hbar}{m}\varphi_t \approx \frac{(x-x_0)^2}{2t}
+\frac{(\bold{r}_{\perp}-\bold{r}_{\perp,1})^2}{2(t-t_c)}
+\frac{\bold{r}_{\perp,1}^2}{2t_c}\ .
\end{equation}
Then, by (\ref{AFraAp}) and (\ref{Fra2dAp}), we get the Fraunhofer
approximation to second order for the one-point source propagator:
\begin{multline}\label{KSFraAp}
K_t(\bold{r},\bold{r}_0|\bold{r}_1)\approx\\
\sigma_{t,t_c}(x,x_0)\frac{e^{im\frac{(x-x_0)^2}{2\hbar t}}}{\sqrt{2i\pi\hbar t/m}}
\frac{e^{im\frac{(y-y_1)^2}{2\hbar(t-t_c)}}}{\sqrt{2i\pi\hbar (t-t_c)/m}}\frac{e^{im\frac{y_1^2}{2\hbar t_c}}}{\sqrt{2i\pi\hbar t_c/m}}
\frac{e^{im\frac{(z-z_1)^2}{2\hbar(t-t_c)}}}{\sqrt{2i\pi\hbar (t-t_c)/m}}\frac{e^{im\frac{z_1^2}{2\hbar t_c}}}{\sqrt{2i\pi\hbar t_c/m}}\ .
\end{multline}
Consequently, for the single-slit model, by integrating over
$y_1,z_1$ on $[-b,b]\times[-a,a]$, we get the usual truncation
approximation formula (\ref{Ktr1}), see Appendix 1, multiplied by a
constant factor $\sigma_{t,t_c}$ depending on $t$ and $t_c$, as
well as on $|x_0|$, $x$ and on the boundary conditions. In
addition, by the semi-classical probabilistic interpretation, see
(\ref{Psclim}), since $\sigma_{t,t_c}(x,x_0)$ is a constant
number, we get the same probability density formula as the one in
\cite{Beau}, (see also (\ref{Ptr}) in Appendix 1) which means that
the initial boundary conditions on the slit do not affect the
diffraction pattern in this regime. We observe this phenomenon
numerically in Fig.3. for $t=0.05$ and $t=0.005$.

\subsection{The fourth-order approximation in the Fraunhofer regime}

Remember that the Fresnel numbers (see Appendix 1) are given by
$$N_F(a)=\frac{2a^2}{\lambda L},\ N_F(b)=\frac{2b^2}{\lambda L}$$ where $L=|x|$ and $\lambda=2\pi
\hbar/(mv)$, and where $a$ is the dimension of the slit along the $z$-axis and $b$ the one along the $y$-axis,
with $v\approx v_x=|x-x_0|/t$. 
In the Fraunhofer regime, we have $N_F(a)\ll 1$ and since the distance between two
successive minima on the pattern is $\Delta z\sim \lambda
L/(2a)=a/N_f(a)$ (see Fig. 3 for the truncation model (TM) and
\cite{Beau}), we have $\Delta z\gg a$ and so we are looking for
the correction for $z\gg a$. We also assume 
that $x\gg b\gg a$ and $2b^2/(\lambda L)\ll1$, so that $\Delta z\gg\Delta y\gg b$. 
In this case, we can neglect the terms of the order
$O(y^4)$ and $O(y_1^4)$ in (\ref{FraAp3}):
\begin{equation}\label{FraAp4}
\frac{\hbar}{m}\varphi_t \approx \frac{x^2}{2t}+\frac{(\bold{r}_{\perp}-\bold{r}_{\perp,1})^2}{2(t-t_c)}
+\frac{\bold{r}_{\perp,1}^2}{2t_c}
-\frac{|x||x_0|}{8t}\left(\frac{(z-z_1)^2}{x^2}
-\frac{z_1^2}{x_0^2}\right)^2
\end{equation}
given that $z\gg a\geq z_1$, we get
\begin{equation}\label{Ap0}
\left(\frac{(z-z_1)^2}{x^2}
-\frac{z_1^2}{x_0^2}\right)^2
\approx\frac{(z^2-2 z_1 z)^2}{x^4}
\end{equation}
which means that we keep only the terms of the order $O(z^2z_1^2)$
(plus the terms of the order $z^4$) and we neglect the terms of the order $O(z_1^4)$.
Inserting the approximation (\ref{Ap0}) in (\ref{FraAp4}), we obtain:
\begin{equation}\label{FraAp5}
\frac{\hbar}{m}\varphi_t \approx \frac{x^2}{2t}+\frac{(y-y_1)^2}{2(t-t_c)}+\frac{y_1^2}{2t_c}
+\frac{(z-z_1)^2}{2(t-t_c)}+\frac{z_1^2}{2t_c}
-\frac{|x_0|z^2}{8|x|^3t}(2z_1-z)^2\ ,
\end{equation}
which leads to a similar expression to the second order expanding
of the phase (\ref{Fra2dAp}) after rewriting the last three terms
of (\ref{FraAp5}) in another explicit form:
\begin{equation}\label{FraAp6}
\frac{\hbar}{m}\varphi_t \approx \frac{x^2}{2t}
+\frac{(y-y_1)^2}{2(t-t_c)}+\frac{y_1^2}{2t_c}
+\frac{(z-z_1^{'})^2}{2(t^{'}-t_c^{'})}+\frac{z^{'2}_1}{2t_c^{'}}-
\frac{z^4}{8|x|^2}\frac{t_c}{t(t-t_c)}\ ,
\end{equation}
where we used the expression (\ref{tcTrc}), and where
\begin{gather}
t_c^{'}=t_c\left(1-\frac{t_c}{2t}\frac{z^2}{x^2}\right)^2\left(1+\frac{4t_c^3}{(t-t_c)^2 t}\frac{z^4}{x^4}\right)^{-1}
\approx t_c\left(1-\frac{z^2}{|x|^2}\frac{t_c}{t}\right)\ ,\label{Prtc}\\
t'-t_c^{'}=t-t_c\ ,\label{Prt}\\
z_1^{'}=z_1\left(1-\frac{z^2}{2|x|^2}\frac{t_c}{t}\right)\ .
\label{Prz1}
\end{gather}
Here we approximated (\ref{Prtc}) to the second order
$O(\frac{z}{x})^2$ such that $\frac{z_1'^2}{2t_1'}$ is of the
fourth order $O(\frac{z}{x})^4$. We remark that the term $\theta
\frac{z^2}{2t},\
\theta\equiv-\frac{z^2}{4|x|^2}\frac{t_c}{(t-t_c)}$ appearing in
(\ref{FraAp6}) gives no contribution to the intensity. Keeping the
zeroth-order approximation for the amplitude (\ref{AFraAp}), the
fourth -order approximation gives the propagator in the
$z$-direction :
\begin{multline}
e^{im\theta \frac{z^2}{2\hbar t}}\int_{-a}^{a}dz_1
\frac{e^{im\frac{x^2}{2\hbar t}}}{\sqrt{2i\pi\hbar t/m}}
\frac{e^{im\frac{(z-z_1^{'})^2}{2\hbar(t^{'}-t_c^{'})}}}{\sqrt{2i\pi\hbar (t-t_c)/m}}
\frac{e^{im\frac{z_1^{'2}}{2\hbar t_c}}}{\sqrt{2i\pi\hbar t_c/m}}\\
=e^{im\theta \frac{z^2}{2\hbar t}}\int_{-a'}^{a'}dz_1^{'}
\frac{e^{im\frac{x^2}{2\hbar t}}}{\sqrt{2i\pi\hbar t/m}}
\frac{e^{im\frac{(z-z_1^{'})^2}{2\hbar(t^{'}-t_c^{'})}}}{\sqrt{2i\pi\hbar (t^{'}-t_c^{'})/m}}
\frac{e^{im\frac{z_1^{'2}}{2\hbar t_c^{'}}}}{\sqrt{2i\pi\hbar t_c^{'}/m}}
\end{multline}
since $dz_1^{'}/\sqrt{t_c^{'}}\approx dz_1/\sqrt{t_c}$ by
(\ref{Prtc}) and since $\sqrt{1+\epsilon^2}\approx
1+\frac{\epsilon^2}{2},\ \mathrm{for}\ \epsilon\ll1$. Thus we
obtain a similar result to (\ref{Ktr1}):
\begin{equation}\label{K4Ap1}
\widetilde{K}_{t,t_c}(x,y,z;b,a)=\sigma_{t,t_c}(x,x_0)\frac{e^{im\frac{(x^2+y^2+z^2)}{2\hbar
t}}}{(2i\pi\hbar t/m)^{3/2}} F_{t',t_c^{'}}(z,a')F_{t,t_c}(y,b),
\end{equation}
where the function $F_{t',t_c^{'}}$ is defined by (\ref{Ktr2}),
and where $\alpha_{t',t_c^{'}}$ is given by
\begin{equation}\label{K4Ap2}
\alpha_{t',t_c^{'}}(z,a^{'})= \left(\frac{2a^{'2}}{2\pi\hbar t_c(t-t_c)/mt}\right)^{1/2}\left(1-\frac{z}{a^{'}}\frac{t_c}{t}\right)
=\alpha_{t,t_c}(z,a^{'})\ .
\end{equation}
Similar to (\ref{Ktr4}), since $\lambda\approx2\pi\hbar/mv_x$ with
$v_x=|x-x_0|/t$ and $t_c\approx |x_0|/v_x=|x_0|t/|x-x_0|$, we can
rewrite (\ref{K4Ap2}) as
\begin{equation}\label{K4Ap3}
\alpha_{t',t_c^{'}}(z,a^{'})\equiv \sqrt{\gamma N_F^{'}}\left(1-\frac{z}{a^{'}\gamma}\right),\ \gamma=|x-x_0|/|x_0|
\end{equation}
where the Fresnel number is:
\begin{equation}\label{NFpr}
N_F^{'}\equiv \frac{2a^{'2}}{\lambda L}\approx N_F\times\left(1-\frac{z^2}{\gamma L^2}\right)\ ,
\end{equation}
where $L=|x|$ and $\gamma=|x-x_0|/|x_0|$ (and we have used
$t_c/t\approx |x_0|/|x-x_0|$). By (\ref{K4Ap1}), we find an
analogous result for the distance $\Delta z$ between two
successive minima of the intensity (see \cite{Beau} for a detailed
approximation)  but to fourth-order approximation:
\begin{equation}\label{DzPr}
\Delta z'\approx \frac{\lambda L}{2a'}\approx\Delta z\left(1+\frac{z^2}{2\gamma L^2}\right)\ .
\end{equation}
Thus, we should observe in the intensity pattern (see Fig.2c), a
deviation of the distance between the consecutive minima from the
truncation approximation given by the following law:
\begin{equation}\label{delta}
\delta \equiv \frac{\Delta z'-\Delta z}{\Delta z}\approx
\frac{z^2}{2\gamma L^2}.
\end{equation}

\subsection{Criterion for the validity of the fourth-order approximation}

As we saw in the last Section IV.B, the 4th-order correction
involves the ratio between the distance of the observation on the
screen and the distance in the $x$-direction between the screen
and the slit. The aim of this section is to understand the
condition of validity for the 4th -order correction to the
truncation approximation model. Here we have to assume that the
system is in the semi-classical regime so that the parameter $\mu$,
introduced in Section III.B is relatively small. Hence we will see
that the corrections are not related to the value of the parameter
$\mu$ but to the quantum fluctuation along the $x$-axis.

We remark that in Fig.3, Appendix 3 the approximation seems valid for 
the curves at the right side (Fig.3.1c-3.3c compared with Fig.3.4c) 
and at the middle (Fig.3.1b-3.3b compared with Fig.3.4b), 
whereas it is not correct for the curves at the left side (Fig.3.1a-3.3a compared with Fig.3.4a). 
Actually, the optical resolution of the pattern has to be
high enough to distinguish two successive minima, i.e. the
relative difference $\delta$, see (\ref{delta}), has to be small :
\begin{equation}\label{C1}
\delta\lll 1 \Rightarrow z\ll L\sqrt{2\gamma}
\end{equation}
This means that the distance $z$ on the screen has to be
relatively small compared to $L$ (if $\gamma\sim 1/2$). We have
seen that in the Fraunhofer regime (see Appendix 1 and \cite{Beau}
for more details) the pattern on the screen has several minima at
$n\Delta z$, $n=2,3,4,\cdots$ with distances $\Delta z=\lambda
L/2a$. In fact, for $z\gg\Delta z$, the intensity becomes rather
small compared to its maximum and so the visibility is low. If we
consider that we can observe the pattern in a window $|z|\leq
n\Delta z$, then by (\ref{C1}) we get the criterion
\begin{equation}\label{C2}
\Delta z\ll L\frac{\sqrt{2\gamma}}{n} \Rightarrow \frac{\lambda}{L}\ll \frac{a}{L}\frac{\sqrt{8\gamma}}{n}
\end{equation}
Hence, given the geometrical parameter $a/L$, we obtain a
condition for the ratio between the wavelength $\lambda=
2\pi\hbar/mv\approx2\pi\hbar t/(m|x-x_0|)$ and the distance $L$.
We reintroduce the parameter $\lambda_0=\left(\frac{2\pi\hbar t}{m}\right)^{1/2}$, 
so that $\lambda=\frac{\lambda_0^2}{|x-x_0|}$. The length
$\lambda_0$ can be interpreted as the spatial fluctuation of the
phase $\exp(im|x-x_0|^2/2\hbar t)=\exp{(i\pi
|x-x_0|^2/\lambda_0^2)}$ appearing in the free propagator
(\ref{green0}). So if $t$ is small enough, then
$\lambda/|x-x_0|=\lambda_0^2/|x-x_0|^2$ will be very small and
consequently the space fluctuations in the $x$-axis will be
negligible. On the contrary, if $t$ is large, the fluctuations are
not negligible and then the approximation $\lambda\ll |x-x_0|$ is not
valid. The criterion (\ref{C2}) gives a condition:
\begin{equation}\label{C3}
\frac{\lambda_0^2}{|x-x_0|^2}\ll \frac{1}{n}\frac{a}{L}\left(\frac{8\gamma}{\gamma'}\right)^{1/2} \Leftrightarrow q\ll\frac{1}{n}
\end{equation}
where the parameter called the \textit{coherence number in the $x$-direction} $q$ is defined as
\begin{equation}\label{q}
q\equiv \frac{\kappa^2}{\rho}
\end{equation}
where $\kappa\equiv\lambda_0/|x-x_0|$ is the \textit{quantum fluctuation parameter in the $x$-axis}
and $\rho\equiv\frac{a}{L}\left(\frac{8\gamma}{\gamma'}\right)^{1/2}$ with $\gamma'=|x-x_0|/L$, is the inverse of the \textit{zoom parameter}.

In the Fig.3, Appendix 3 
for the diffraction patterns at the left side (Fig.3.1a-3.4a) 
the coherence parameter is of the order of unity, 
and the semi-classical parameter $\mu\sim 2500$ 
(with $N_F(a)\sim 3 \times 10^{-5}$, $N_F(b)\sim 3 \times 10^{-3}$).
We observe some differences with the truncation approximation.
First, the distance between the fringes is about $66\%$ in case of the first minimum 
and $100\%$ for the second.
We conclude that in those cases, the fourth-order approximation above is not valid. 
However, we observe different shapes for the different boundary conditions
and also the amplitude of the oscillations are not the same. 
If we suppose that we can experimentally (and this is probably a big challenge)
build an apparatus for which the parameter $\mu$ is of the range $10^2-10^{4}$,
the differences described before would provide physical information about the surface of the slit. 

For the Fig.3.1b-3.4b, the parameters $q\sim 6\times 10^{-2}$ and $\mu\sim 5\times 10^4$
(with $N_F(a)\sim 6 \times 10^{-4}$, $N_F(b)\sim 3 \times 10^{-2}$).
In this case the fourth-order approximation is quite good for the first ten fringes.
For example, we have $\delta\sim 0.1\%$ for the first fringe' shift, $\delta\sim 7 \%$ for the third one 
and $\delta\sim 16\%$ for the fifth one.
Additionally, we observe that the shape of the curves as well as the amplitude of the oscillations 
do not depend on the boundary conditions.

Similarly for the Fig.3.1c-3.4c, the parameters $q\sim 6\times 10^{-3}$ and $\mu\sim 5\times 10^{5}$ 
(with $N_F(a)\sim 6 \times 10^{-3}$, $N_F(b)\sim 6 \times 10^{-1}$)
and we obtain diffraction patterns very close to the truncation approximation.
In the next section, we will see that experimentally this is the general situation.

\section{Discussion}

\subsection{General remarks}

Concerning the two-slit problem, we observed that for large $\mu$ and for small Fresnel numbers 
($N_F(a),N_F(b),N_F(d)\ll1$, where $d$ is the distance between the centers of the slits along the $z$-axis)
the interference pattern has a similar shape to the one for the truncation approximation (see \cite{Beau}), see Fig. 4, Appendix 3
but we observe a similar diffraction in time phenomenon for the envelope of the interference pattern
which is nothing but the diffraction curve for a single-slit centered at $x_1=y_1=z_1=0$.
The two-slit propagator formula is given by:
\begin{equation}\label{Kdble}
K^{(dble)}(\bold{r},t,\bold{r}_0,0)=
\int_{-a-d}^{-a+d}dz_1\int_{-b}^bdy_1\ K(\bold{r},t;\bold{r}_0,0|\bold{r}_1)
+\int_{a-d}^{a+d}dz_1\int_{-b}^bdy_1\ K(\bold{r},t;\bold{r}_0,0|\bold{r}_1)
\end{equation}
where the three-dimensional one-point source propagator $K(\bold{r},t;\bold{r}_0,0|\bold{r}_1)$ is given by (\ref{KSsol}).

Here we have only studied a rectangular aperture but naturally
the propagator (\ref{KSlit1}) can be generalized for an arbitrarily shape of slit:
\begin{equation}
K^{(\Sigma)}(\bold{r},t,\bold{r}_0,0)=\int_{\Sigma}dy_1dz_1\ K(\bold{r},t;\bold{r}_0,0|\bold{r}_1)
\end{equation}
where $\Sigma$ is the aperture of the slit (e.g., circle)
and where the one-point source propagator $K(\bold{r},t;\bold{r}_0,0|\bold{r}_1)$ is given by (\ref{KSsol}). \\

Also, we mention that the semi-classical approximation is valid for two dimensions
where we have to integrate (\ref{KSlit1}) along the $y_1$-axis 
and apply the stationary phase approximation method.
Similarly to (\ref{SC3}) we get:
\begin{equation}\label{2DSC}
K_{sc}^{(2D)}(\bold{r},t;\bold{r}_0,0)
=\frac{e^{\frac{im x^2}{2\hbar t}}}{\sqrt{2i\pi\hbar t/m}}\int_{-a}^{a}dz_1\ \sigma_{t,\tau_{sc}}^{(2D)}(x,x_0)
\frac{e^{\frac{im (z-z_1)^2}{2\hbar(t-\tau_{sc})}}}{\sqrt{2i\pi\hbar (t-\tau_{sc})/m}}
\frac{e^{\frac{im (z_1-z_0)^2}{2\hbar\tau_{sc}}}}{\sqrt{2i\pi\hbar\tau_{sc}/m}}
\end{equation}
where $\sigma_{t,\tau_{sc}}^{(2D)}(x,x_0)$ is given by (\ref{sigma})
and with the semi-classical time $\tau_{sc}$ is given by (\ref{tc}) taking $y_0=y_1=y=0$.

\subsection{Ultracold atoms slit experiment under gravity}

In the slit experiments for electrons, cold neutrons and heavy molecules, 
we point out that the dimensions of the apparatus are large so that 
the semi-classical parameter is very large, 
for example in \cite{ZeilingerNeutron} and \cite{Bach}, $\mu$ is of the order $\sim 10^{10}$,
it is $\sim10^{13}$ in \cite{ZeilingerC60} and $\sim10^{7}$ in \cite{Takuma}.
Additionally, experimentally the initial wave function (at the time of the emission)
is not localized at $\bold{r}_0$ but a plane wave along the $x$-axis. 
As discussed above, it suffices to Fourier transform the propagator (\ref{KRew})
with respect to the variable $x_0$. 
However, in the cold atoms experiment \cite{Takuma}
the narrow wave packet model is more convenient even if for realistic conditions
we do not reach the limit $\sigma\rightarrow0$. 
In \cite{Gondran} is presented a theoretical description and interpretation of the latter experiment,
but still following the truncation approximation. 
Concretely, a bunch of coherent cold neon atoms (mass $m=3.349\times10^{-26}\mathrm{kg}$) 
is trapped above a plate where there are two apertures (the two-slits system) at a distance $l_1$.
At the time $t_0=0$ the optical-magnetic trap is switch off 
and the atoms fall under the gravity field of the earth,
passing through the two slits and strike a detection plate at distance $l_2$ from the two-slit plate.
It is assumed that the initial wave of an atom falling down the gravity field
is a Gaussian wave packet centered at $\bold{r}_0=\bold{0}$ with an initial vector wave $\bold{k}_0$.
It is also assumed that the motion along the $z$-axis is classical. 
Moreover, the dimensions of the slits are considered to be small compared to the distances $l_1$ and $l_2$ 
which, in addition to the classical treatment along $z$, is the truncation approximation.
By fixing $k_{0,z}=0$, it follows that the classical time that the particle needs to pass through the slits
is simply given by $t_1=\sqrt{2l_1/g}$. 
Then, taking $l_1\sim 0.1\mathrm{m}$, $g=9.81\mathrm{ms}^{-2}$ we compute $t_1\sim 0.1\mathrm{s}$
and so $\lambda\approx \hbar/mv_z\sim 1.5\times 10^{-8}\mathrm{m}$
which yields to $\mu\approx 2\pi (l_1+l_2)/\lambda\sim 10^{7}$.
Besides, in \cite{Gondran} numerical simulations are performed to study
the probability density for small $l_2$
showing that the interference patterns can be observed only for $l_2\geq 5\times 10^{-4}\mathrm{m}$.
In the Fig. 4 of \cite{Gondran}, the probability density is plotted especially 
for $l_1=10^{-4}\mathrm{m},\ 5\times 10^{-4}\mathrm{m},\ 10^{-3}\mathrm{m},\ 1.13\times10^{-1}\mathrm{m}$
where we can observe a transition between the so-called \textit{separated regime} and the \textit{mixed regime} \cite{Beau} 
between the two-slits patterns.  
However the length $l_1$ was kept fixed for numerical simulation and so we may wonder what 
the shape pattern would be if we varied the length $l_1$ in such a way that the total length $l_1+l_2$ 
is of orders $10^{-6}\ -\ 10^{-2}\mathrm{m}$.
In this situation, the semi-classical parameter $\mu$ would be of orders $10^{2}\ -\ 10^6$
and so we expect the probability density to be modified by the fluctuation along the $z$-axis
which should exhibit a correction to the truncation approximation model.

Here we will give a brief description of the modifications needed in our model to describe this experiment. 
Again we use the Brunker-Zeilinger method 
and obtain a semi-classical approximation for the propagator. 
We leave the numerical simulation for another article where we will investigate the phenomenological consequences of our model.

As described above, we consider Gaussian wave packet (\ref{Psi0}) centered at $x_0=y_0=z_0=0$
falling down the gravity field $\bold{g}=g\ \bold{e}_z$ where $g=9.81\ \mathrm{ms}^{-1}$ 
and passing through a single slit, where the slit is an aperture in a plane orthogonal to the $z$-axis 
positioned at $z_1>0$. After a time $t$, the particle is detected on a screen at the position $z>z_1$.
We describe the quantum-motion of the particle by the following equation similar to (\ref{SchEq}):
\begin{equation}\label{SchEqG}
\begin{array}{ll}
-\frac{\hbar^2}{2m}\nabla^2 \psi(\bold{r},t) + V(z)\ \psi(\bold{r},t) = i\hbar\frac{\partial}{\partial t}\psi(\bold{r},t)\\
\psi(\bold{r},t) = 0 \mbox{ for } z
> z_1 \mbox{ and } t< t_1, \mbox{ and }
\psi(\bold{r}_1,t)=\phi(\bold{r}_1,t) \mbox{ for } t > t_1.
\end{array}
\end{equation}
with $V(z)=mgz$ and where we fixed the boundary and initial condition on the plane of the slit. 
As before, we consider a shutter opening at the time $t_1=0$ 
after which the wave propagates below the slit plane (i.e. $z>z_1$).   

Using the same arguments as previously but replacing the free Green's function by:
\begin{equation}
G_g(\bold{r},t;\bold{r}',t')=G_0(\bold{r},t;\bold{r}',t')\ e^{\frac{im}{2\hbar}\left(g(z+z')(t-t')-\frac{g^2}{12}(t-t')^3\right)} \\
\end{equation}
where the free propagator $G_0(\bold{r},t;\bold{r}',t')$ is given by (\ref{green0}), we get:
\begin{equation}
K^{(g)}(\bold{r},t;\bold{0},0)
=\frac{i\hbar}{2m}\int_{0}^{t}d\tau \int_{-a}^a dz_1\int_{-b}^b dy_1\ \chi_{t,\tau}(z,z_1)
G_g(\bold{r},t;\bold{r}_1,\tau)G_g(\bold{r}_1,t;\bold{0},\tau)
\end{equation}
where:
$$\chi_{t,\tau}(z,z_1)\equiv
\eta_1\frac{z_1}{\tau}+\eta_2\frac{z-z_1}{t-\tau}-i\eta_2\ gt+i(\eta_1+\eta_2)g\tau-i(\eta_1-\eta_2)\frac{g\tau}{2}$$

Then for $\mu\equiv m\bold{r}^2/(2\hbar t)\gg1$, by the stationary phase approximation we obtain:
\begin{equation}
K_{sc}^{(g)}(\bold{r},t;\bold{0},0)
\approx \int_{-a}^{a} dx_1\int_{-b}^b dy_1\ A_{sc}(\bold{r},t;\bold{0},0|\bold{r}_1)\ e^{i\phi_{sc}(\bold{r},t;\bold{0},0)|\bold{r}_1)} 
\end{equation}
where the one-point source amplitude is given by
$$A_{sc}(\bold{r},t;\bold{0},0|\bold{r}_1)
=\frac{\chi_{t,\tau_{sc}}(z,z_1)}{((2i\pi\hbar/m)^2(t-\tau_{sc})\tau_{sc})^{3/2}}
\left(\frac{2i\pi}{\omega_{sc}(\bold{r},t;\bold{0},0)}\right)^{1/2}$$
with 
$$\omega_{sc}(\bold{r},t;\bold{0},0|\bold{r}_1)
=\frac{m}{\hbar}\left(\frac{(\bold{r}-\bold{r}_1)^2}{(t-\tau_{sc})^3}+\frac{|\bold{r}_1|^2}{(\tau_{sc}-t_0)^3}\right)
-\frac{mg^2t}{4\hbar}$$
and the one-point source phase by 
\begin{multline*}
\phi_{sc}(\bold{r},t;\bold{0},0|\bold{r}_1)=
\frac{m}{2\hbar}\left(\frac{(\bold{r}-\bold{r}_1)^2}{t-\tau_{sc}}+\frac{|\bold{r}_1|^2}{\tau_{sc}-t_0}\right)\\
+\frac{m}{2\hbar}\left(g(z+z_1)(t-\tau_{sc})+(z_1)(\tau_{sc}-t_0)
-\frac{g^2}{12}(t-\tau_{sc})^3-\frac{g^2}{12}(\tau_{sc}-t_0)^3\right)
\end{multline*}
Here, the semi-classical time $\tau_{sc}$ is the solution of the following fifth-order polynomial equation 
(since after expansion, the term in $\tau^6$ disappears):
\begin{equation}\label{tcg}
(\bold{r}-\bold{r}_1)^2\tau^2-|\bold{r}_1|^2(t-\tau)^2=gz(t-\tau)^2\tau^2
+\frac{g^2}{4}\tau^4(t-\tau)^2-\frac{g^2}{4}\tau^2(t-\tau)^4
\end{equation}
which can be interpreted as the conservation of the classical energy of the particle passing through the aperture,
the trajectories being two broken parabolas similarly to the broken straight lines for the case without gravity. 
The first parabolic trajectory goes from $\bold{r}_0=\bold{0}$ at the time $t_0=0$
to the point $\bold{r}_1$ at the time $\tau$, 
and the second one from $\bold{r}_1$ at the time $\tau$
to the point $\bold{r}$ at the time $t$.
Then, the classical energies for a particle following the two trajectories are 
$E_0=\frac{m}{2}\bold{v}_0^2$  for the first one and
$E_1=\frac{m}{2}\bold{v}_1^2-mgz_1$ for the second one,  
with the classical velocities 
$$\bold{v}_0=\frac{x_1-x_0}{\tau-t_0}\bold{e}_x+\frac{y_1-y_0}{\tau-t_0}\bold{e}_y+(\frac{z_1}{\tau-t_0}-\frac{g}{2}(\tau-t_0))\bold{e}_z$$
$$\bold{v}_0=\frac{x-x_1}{t-\tau}\bold{e}_x+\frac{y-y_1}{t-\tau}\bold{e}_y+(\frac{z-z_1}{t-\tau}-\frac{g}{2}(t-\tau))\bold{e}_z$$
The conservation of energy equation $$E_0=E_1 \Leftrightarrow \frac{1}{2}|\bold{v}_0|^2=\frac{1}{2}|\bold{v}_1|^2-gz_1$$ 
then leads to the equation (\ref{tcg}).\\

\subsection{Conclusion}

To summarize, the fourth-order corrections are generally small for realistic experimental situations,
which is to be expected a priori since the theoretical predictions fit very well with the past and current experiments.
However, our model brings a new perspective investigating the quantum diffraction beyond the truncation approximation.
The fluctuation along the ``propagation axis'' could in principle be quantitatively and qualitatively demonstrated experimentally 
for a system following the conditions (minimum):
\begin{enumerate}
 \item the apparatus has to be of mesoscopic scale, 
since we have seen that the length along the ``propagation axis'' has to be of orders $10^{-6}-10^{-3} \mathrm{m}$
to have $\mu\sim 10^{2}-10^5$ in order to be able to observe any shift in the distances between the fringes.
 \item  the statistics have to be high enough to have a good accuracy so that we can detect a shift for $\delta\sim 10\%$
and also so that the differences between the amplitude of the oscillation for different boundary conditions can be detected.
\end{enumerate}
To construct an experimental apparatus of this kind is certainly a challenge 
but perhaps not entirely beyond future advances in technology.

\newpage

\section{Appendices}

\subsection{Appendix 1: The truncation approximation for the quantum-multi-slit diffraction problem}

We recall\cite{Beau} that the single ``gate'' slit propagator,
centered on the $x$-axis of size $2a$ on the $z$-axis and $2b$ on
the $y$-axis is given by the following formula \cite{Beau}:
\begin{equation}\label{Ktr}
K_{\textrm{Trunc}}^{(a,b)}(\bold{r},t;\bold{r}_0,0|\bold{r}_1,t_c)=\frac{e^{i\frac{m(x-x_0)^2}{2\hbar t}}}{(2i\pi\hbar t/m)^{1/2}}
\int_{-a}^{a}dz_1 \int_{-b}^{b}dy_1 \frac{e^{i\frac{m[(y-y_1)^2+(z-z_1)^2]}{2\hbar(t-t_c)}}}{2i\pi\hbar(t-t_c)/m}
\frac{e^{i\frac{m[(y_1-y_0)^2+(z_1-z_0)^2]}{2\hbar t_c}}}{2i\pi\hbar t_c/m}\ ,
\end{equation}
where $t_c = |x_1-x_0|t/|x-x_0|$. We note that this formula is
valid only if the distances on the $x$-axis are very large
compared to the distance on the screen and to the sizes of the
slit, i.e. for $|x-x_1|,|x_1-x_0|\gg a,b,|y|,|z|$.\cite{Beau} The
formula (\ref{Ktr}) can be interpreted as follows: The propagator
is the sum over the paths $x(\tau)$ of the particle going from the
source $(x(0)=x_0,z(0)=z_0,y(0)=y_0)$ to the screen
$(x(t)=x,z(t)=z,y(t)=y)$, given that it goes through the slit
$x(t_c)=x_1$, $z(t_c)=z_1\in[-a,a],\ y(t_c)=y_1\in[-b,b]$ at the
time $t_c$ defined just above. Notice that we can find an explicit
formula for the propagator (\ref{Ktr}) in terms of the Fresnel
function \cite{FH}, \cite{Beau}:
\begin{equation}\label{Ktr1}
K_{\textrm{Trunc}}^{(a,b)}(\bold{r},t;\bold{r}_0,0|\bold{r}_1,t_c)=
\frac{e^{i\frac{|\bold{r}-\bold{r}_0|^2}{2t}}}{(2i\pi\hbar t/m)^{3/2}} F_{t,t_c}(z,a)F_{t,t_c}(y,b)
\end{equation}
where:
\begin{equation}\label{Ktr2}
F_{t,t_c}(z,a)\equiv
\left(C[\alpha_{t,t_c}(z,a)]+C[\alpha_{t,t_c}(z,-a)]+iS[\alpha_{t,t_c}(z,a)]+iS[\alpha_{t,t_c}(z,-a)]\right)
\end{equation}
and where the Fresnel functions are defined as follows \cite{Abramowitz}:
\begin{gather}
u\in\mathbb{R}^1\mapsto C[u]=\int_{0}^{u}dw\ \cos{(\frac{\pi w^2}{2})}\label{C}\\
u\in\mathbb{R}^1\mapsto S[u]=\int_{0}^{u}dw\ \sin{(\frac{\pi w^2}{2})}\label{S}
\end{gather}
with
\begin{equation}\label{Ktr3}
\alpha_{t,t_c}(z,a)\equiv \left(\frac{ma^2t}{\pi\hbar t_c(t-t_c)}\right)^{1/2}\left(1-\frac{z}{a}\frac{t_c}{t}\right)
\end{equation}
and since we have seen that $\lambda\approx2\pi\hbar/mv_x$ with $v_x=|x-x_0|/t$ and $t_c\approx |x_1-x_0|/v_x=|x_1-x_0|t/|x-x_0|$,
we can rewrite (\ref{Ktr3}) as
\begin{equation}\label{Ktr4}
\alpha_{t,t_c}(z,a)\equiv \sqrt{\gamma N_F}\left(1-\frac{z}{a\gamma}\right)\ ,
\end{equation}
where $\gamma=|x-x_0|/|x_1-x_0|$ and where the Fresnel number is defined as
\begin{equation}\label{NF}
N_F\equiv \frac{2a^2}{\lambda L}\ ,
\end{equation}
where $L=|x-x_1|$.

The intensity on the screen is proportional to the modulus square
of the propagator, and  is given by\cite{Beau}:
\begin{equation}
I_{\textrm{Trunc}}^{(a,b)}(\bold{r},t;\bold{r}_0,0|\bold{r}_1,t_c)|
\equiv I_0 |K_{\textrm{Trunc}}^{(a,b)}(\bold{r},t;\bold{r}_0,0|\bold{r}_1,t_c)|^2
=I_0\left(\frac{m}{2\pi\hbar t}\right)^3|F_{t,t_c}(z,a)|^2|F_{t,t_c}(y,b)|^2\label{Itr}
\end{equation}
In \cite{Beau} three distinct regimes were considered
depending on the value of the Fresnel number: \\
(i) If $N_F\ll1$: we have the \textit{Fraunhofer regime},
which means that the distances are sufficiently large to get a usual interference pattern \cite{Optics}
where the distance between two consecutive minima of intensity are about $\lambda L/(2a)$ in the $z$-direction
and $\lambda L/(2b)$ in the $y$-direction. \\
(ii) If $N_F\gg1$: we are in the so-called \textit{Fresnel regime}
where the interference pattern has a similar shape as the gate but with a different scale: $a(\gamma-1)$ in the $z$-direction
and $b(\gamma-1)$ in the $y$-direction.
More specifically, the intensity is very small if $z>a(\gamma-1)$ and $y>b(\gamma-1)$
and oscillate very fast around a constant if $z<a(\gamma-1)$ and $y<b(\gamma-1)$. \\
(iii) If $N_F\sim1$: the \textit{intermediate regime} is a transition between both regimes
for which there is a spreading around the center of the intensity
and similar to the Fraunhofer regime for large distances (on the screen).  \\

The formula for the multi-slit problem is given by the sum over
the single-slit propagator for each slit (centered in $(A_j,B_j),\
j=1,..,N$):
\begin{equation}\label{KsclN}
K_{\textrm{Trunc}}^{(N)}(\bold{r},t;\bold{r}_0,0|\bold{r}_1,t_c)=\frac{e^{i\frac{m(x-x_0)^2}{2\hbar t}}}{(2i\pi\hbar t/m)^{1/2}}\sum_{j=1}^{N}
\int_{A_j-a_j}^{A_j+a_j}\int_{B_j-b_j}^{B_j+b_j}\frac{e^{i\frac{m[(y-y_1)^2+(z-z_1)^2]}{2\hbar(t-t_c)}}}{2i\pi\hbar(t-t_c)/m}
\frac{e^{i\frac{m[(y_1-y_0)^2+(z_1-z_0)^2]}{2\hbar(t-t_c)}}}{2i\pi\hbar(t-t_c)/m}\ .
\end{equation}
Then, to get the interference pattern on the screen, we have to
compute the square modulus of the $N$-slit propagator:
\begin{equation}
I_{\textrm{Trunc}}^{(N)}(\bold{r},t;\bold{r}_0,0|\bold{r}_1,t_c)\equiv I_0\times|K_{\textrm{Trunc}}^{(N)}(\bold{r},t;\bold{r}_0,0|\bold{r}_1,t_c)|^2\ .
\end{equation}

\textit{Remark}.: In \cite{Beau}, for a initial Gaussian wave packet in the limit $\sigma\rightarrow0$ ($\sigma$ is the width of the Gaussian),
it is proved that the density of probability is proportional to the square of the propagator:
\begin{equation}\label{Ptr}
P^{(N=1)}(\bold{r},t;\bold{r}_0,0|\bold{r}_1,t_c)=\frac{\pi^2\hbar^2 t_c^2}{m ab}\frac{2\pi\hbar t_c}{m} |K^{(N=1)}(\bold{r},t;\bold{r}_0,0|\bold{r}_1,t_c)|^2
=\frac{1}{4ab\gamma^3}|F_{t,t_c}(z,a)|^2|F_{t,t_c}(y,b)|^2\ ,
 \end{equation}
where we generalized the probabilistic interpretation developed in \cite{Beau} for a two dimensional slit,
and where the factor $\frac{2\pi\hbar t_c}{m}$ in the second equality in (\ref{Ptr}) come from the Gaussian normalisation in the $x$-direction.
Notice that by (\ref{KsclN}), we can extend the result for all $N$ by recursion.\\

\subsection{Appendix 2: Derivation of the one-point source propagator}

By (\ref{KS1}), the one-point source propagator is given by
\begin{gather}\label{Ann1}
K(\bold{r},t;\bold{r}_0,0|\bold{r}_1)\equiv
\int_0^t d\tau\left[\frac{-x_0}{\tau}\eta_1+\frac{x}{t-\tau}\eta_2\right]G_0(\bold{r}-\bold{r}_1,t-\tau)G_0(\bold{r}_1-\bold{r}_0,\tau)\\
=\eta_2K^{(D)}(\bold{r},t;\bold{r}_0,0|\bold{r}_1)
+\eta_1K^{(N)}(\bold{r},t;\bold{r}_0,0|\bold{r}_1)
\end{gather}
where we have introduced the Dirichlet part:
\begin{eqnarray}\label{AnnD1}
K^{(D)}(\bold{r},t;\bold{r}_0,0|\bold{r}_1)
=\int_0^t d\tau\left[\frac{x}{t-\tau}\right]G_0(\bold{r}-\bold{r}_1,t-\tau)G_0(\bold{r}_1-\bold{r}_0,\tau)
\end{eqnarray}
and the Neumann part:
\begin{eqnarray}\label{AnnN1}
K^{(N)}(\bold{r},t;\bold{r}_0,0|\bold{r}_1)
=\int_0^t d\tau\left[\frac{-x_0}{\tau}\right]G_0(\bold{r}-\bold{r}_1,t-\tau)G_0(\bold{r}_1-\bold{r}_0,\tau)
\end{eqnarray}

We will use the Laplace transform defined by
\begin{eqnarray*}
LT\left[f(\tau);\tau,s\right]=\int_{0}^{+\infty}d\tau\ e^{-s\tau}f(\tau)
\end{eqnarray*}
and the inverse Laplace transform
\begin{eqnarray*}
LT^{-1}\left[F(s);s,\tau\right]=\int_{c-i\infty}^{c+i\infty}\frac{ds}{2i\pi}\
e^{\tau s}F(s),
\end{eqnarray*}
where $c$ is a real-valued constant chosen such that the integral
remains finite. We have the following Laplace transforms (see
equations (28) P.147 \S 4.5 and (5) P.246 \S 5.6 in
\cite{Handbook}):
\begin{gather}\label{LT}
LT\left[\left(\frac{m}{2i\pi\hbar t}\right)^{3/2}e^{i\frac{m|\bold{r}|^2}{2\hbar t}};t,s\right]=
\frac{e^{i|\bold{r}|\sqrt{2mis/\hbar}}}{2i\pi\hbar|\bold{r}|/m}\\
LT\left[\frac{1}{t}\left(\frac{m}{2i\pi\hbar t}\right)^{3/2}e^{i\frac{m|\bold{r}|^2}{2\hbar t}};t,s\right]=
\frac{e^{i|\bold{r}|\sqrt{2mis/\hbar}}}{2i\pi|\bold{r}|^2}\left(-\sqrt{2mis/\hbar}+\frac{i}{|\bold{r}|}\right)
\end{gather}

So by (\ref{AnnN1}) and (\ref{LT}) we get
\begin{multline}\label{AnnC1}
K^{(N)}(\bold{r},t;\bold{r}_0,0|\bold{r}_1)\\
=\frac{-m x_0}{\hbar}\times LT^{-1}
\left[\frac{e^{i|\bold{r}-\bold{r}_1|\sqrt{2m i s/\hbar}}}{2i\pi|\bold{r}-\bold{r}_1|}
\frac{e^{i|\bold{r}_1-\bold{r}_0|\sqrt{2m i s/\hbar}}}{2i\pi|\bold{r}_1-\bold{r}_0|^2}
\left(-\sqrt{2mis/\hbar}+\frac{i}{|\bold{r}_1-\bold{r}_0|}\right);s,t\right]
\end{multline}
putting $\bold{u}_1=\bold{r}_1-\bold{r}_0$, $\bold{u}_2=\bold{r}-\bold{r}_1$, (\ref{AnnC1}) is equal to
\begin{multline}\label{AnnC2}
\frac{-m x_0}{\hbar} \frac{(|\bold{u}_2|+|\bold{u}_1)|^2}{|\bold{u}_2||\bold{u}_1|^2}\times\\
 LT^{-1}\left[
\frac{e^{i(|\bold{u}_2|+|\bold{u}_1|)\sqrt{2m i s/\hbar}}}{(2i\pi|\bold{u}_2|+|\bold{u}_1|)^2}
\left(-\sqrt{2mis/\hbar}+\frac{i}{|\bold{u}_2|+|\bold{u}_1|}
+\left(\frac{i}{|\bold{u}_1|}-\frac{i}{|\bold{u}_2|+|\bold{u}_1|}\right)\right);s,t\right]
\end{multline}
then using the inversion formulas (\ref{LT}), we get:
\begin{equation}\label{AnnC3}
K^{(N)}(\bold{r},t;\bold{r}_0,0|\bold{r}_1)
=A_t^{(N)}(\bold{r},\bold{r}_0|\bold{r}_1)
e^{i\varphi(\bold{r},\bold{r}_0|\bold{r}_1)}
\end{equation}
where the amplitude is given by:
\begin{equation}\label{AnnC4}
A_t^{(N)}(\bold{r},\bold{r}_0|\bold{r}_1)
= \frac{-x_0}{(2i\pi\hbar t/m)^{3/2}}\left(\frac{m}{2i\pi\hbar t}\frac{(|\bold{r}-\bold{r}_1|+|\bold{r}_1-\bold{r}_0|)^2}
{|\bold{r}-\bold{r}_1||\bold{r}_1-\bold{r}_0|^2}
+\frac{1}{2\pi|\bold{r}_1-\bold{r}_0|^3}\right)
\end{equation}
and the phase by:
\begin{equation}\label{AnnC5}
\varphi_t(\bold{r},\bold{r}_0|\bold{r}_1)
\equiv\frac{m}{2\hbar t}(|\bold{r}-\bold{r}_1|+|\bold{r}_1-\bold{r}_0|)^2
\end{equation}

Similary for the Dirichlet boundary condition, by the symmetries
$t_1\leftrightarrow t-t_1$ and $u_1\leftrightarrow u_2$  in (\ref{AnnC1}),
we get that the amplitude is given by:
\begin{equation}\label{AnnC6}
A_t^{(D)}(\bold{r},\bold{r}_0|\bold{r}_1)
= \frac{x}{(2i\pi\hbar t/m)^{3/2}}\left(\frac{m}{2i\pi\hbar t}\frac{(|\bold{r}-\bold{r}_1|+|\bold{r}_1-\bold{r}_0|)^2}
{|\bold{r}-\bold{r}_1|^2|\bold{r}_1-\bold{r}_0|}
+\frac{1}{2\pi|\bold{r}-\bold{r}_1|^3}\right)
\end{equation}
and that the phase does not change and is given by (\ref{AnnC5}).

\subsection{Appendix 3: Diffraction patterns}

\begin{figure}
\begin{center}
\scalebox{0.55}{\includegraphics{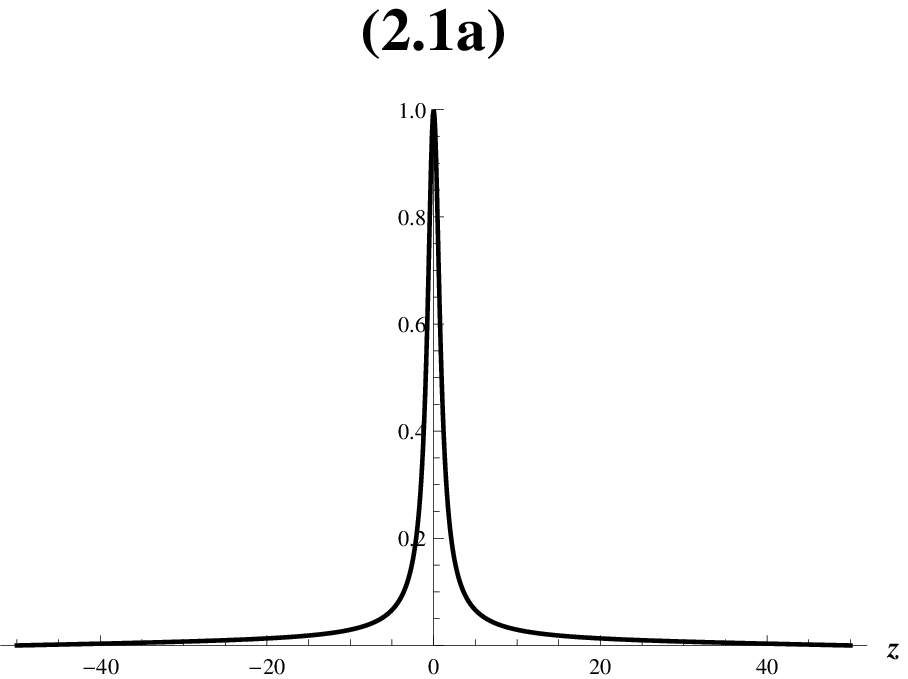}}\scalebox{0.55}{\includegraphics{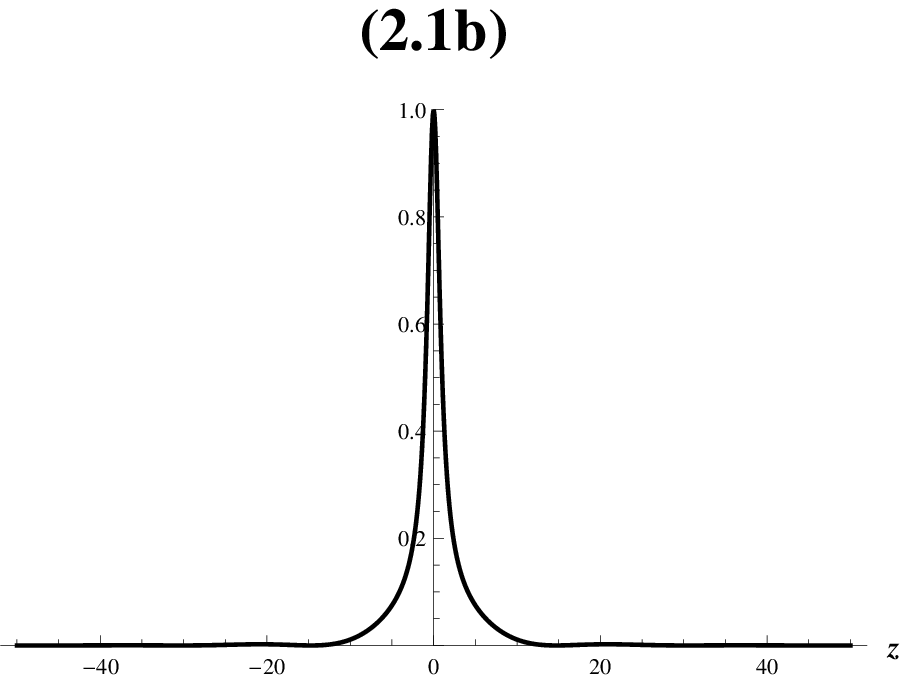}}\scalebox{0.55}{\includegraphics{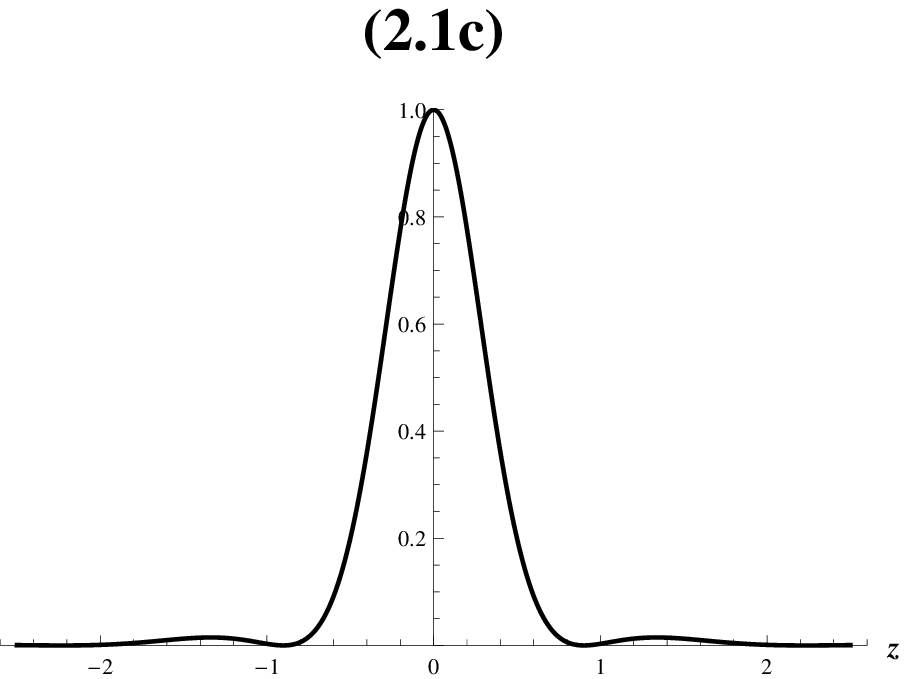}}
\scalebox{0.55}{\includegraphics{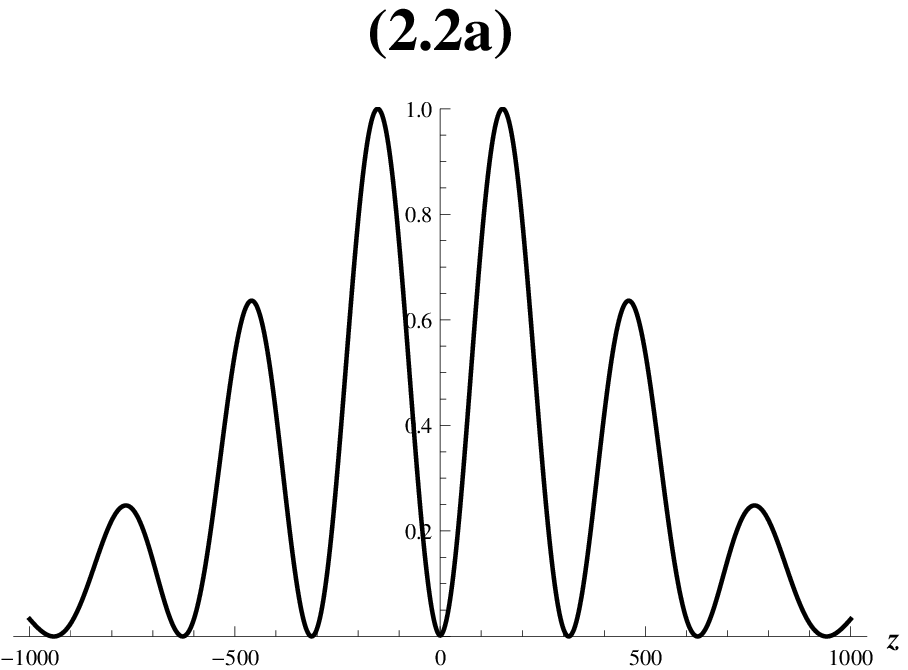}}\scalebox{0.55}{\includegraphics{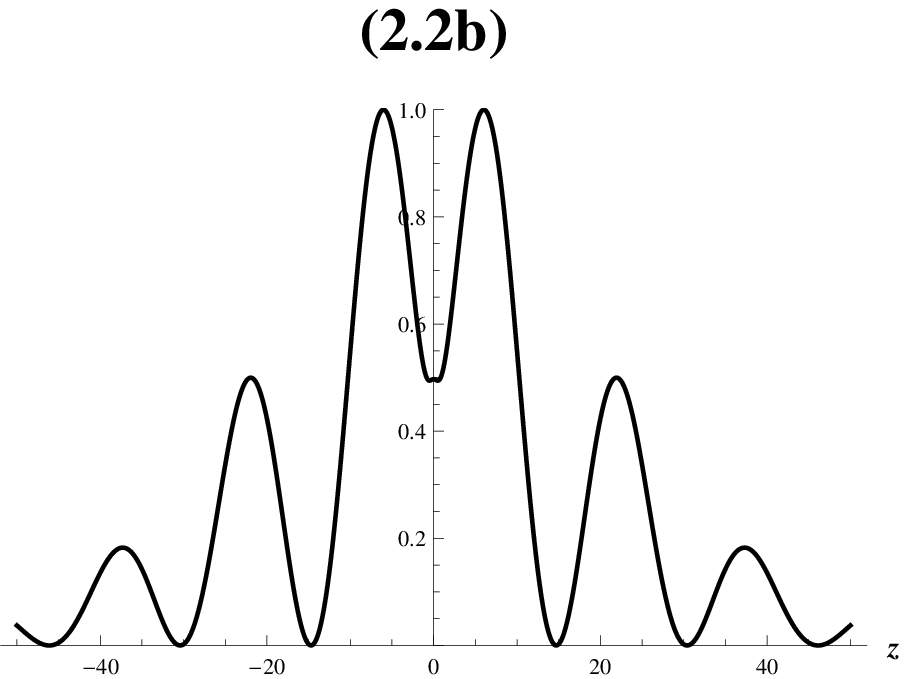}}\scalebox{0.55}{\includegraphics{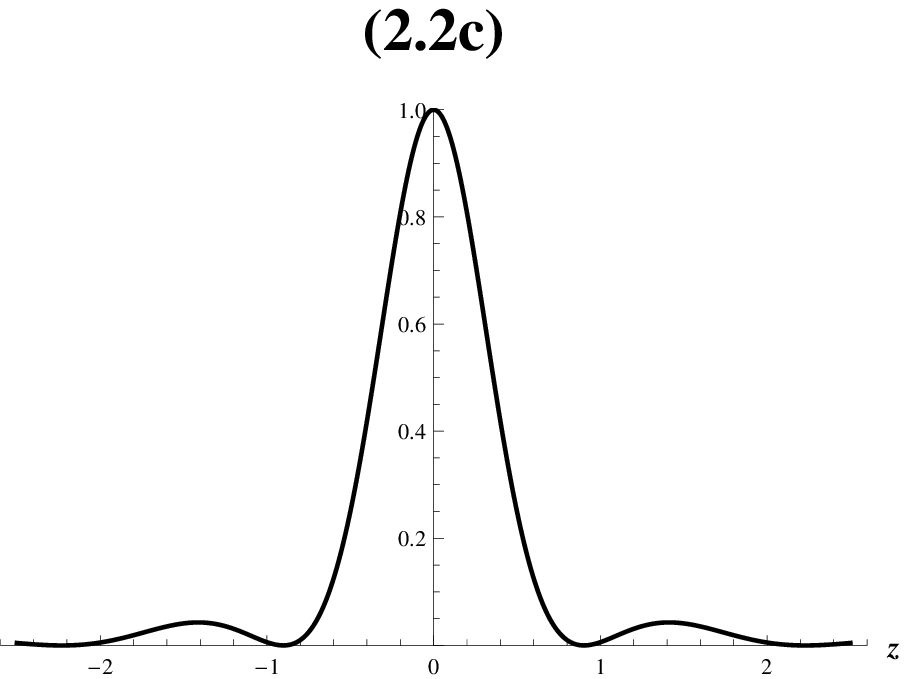}}
\scalebox{0.55}{\includegraphics{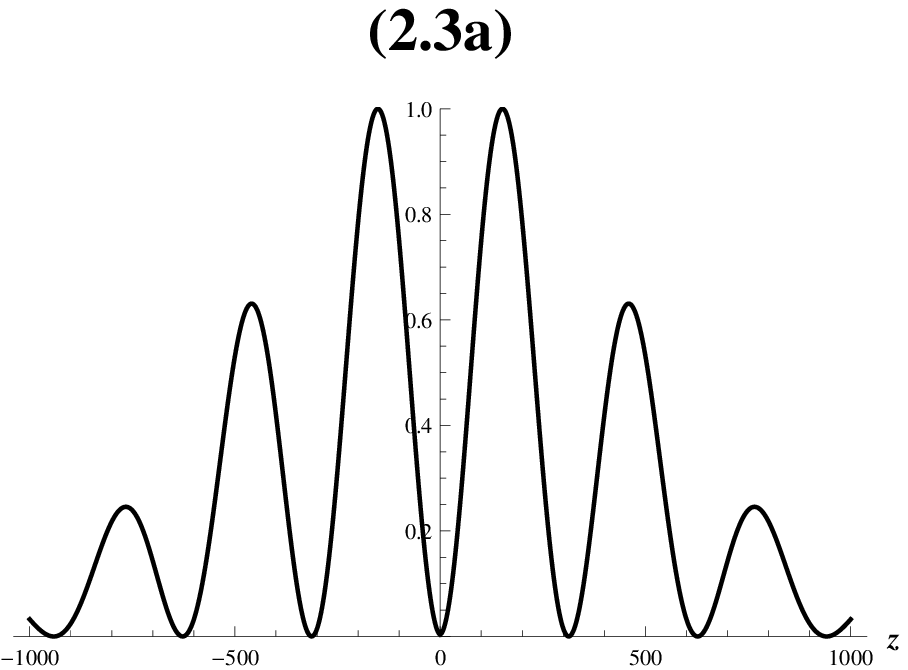}}\scalebox{0.55}{\includegraphics{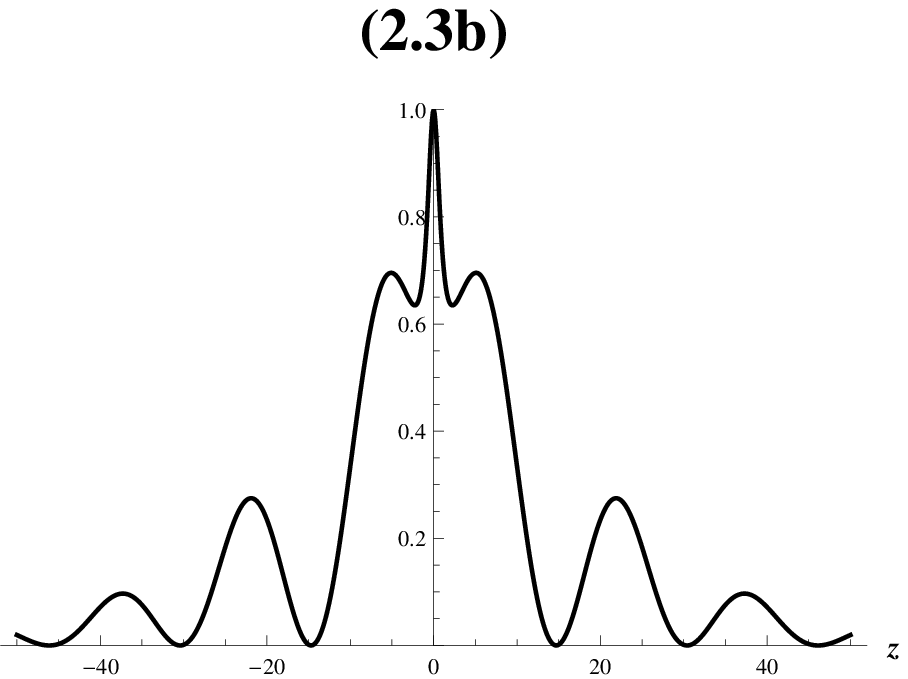}}\scalebox{0.55}{\includegraphics{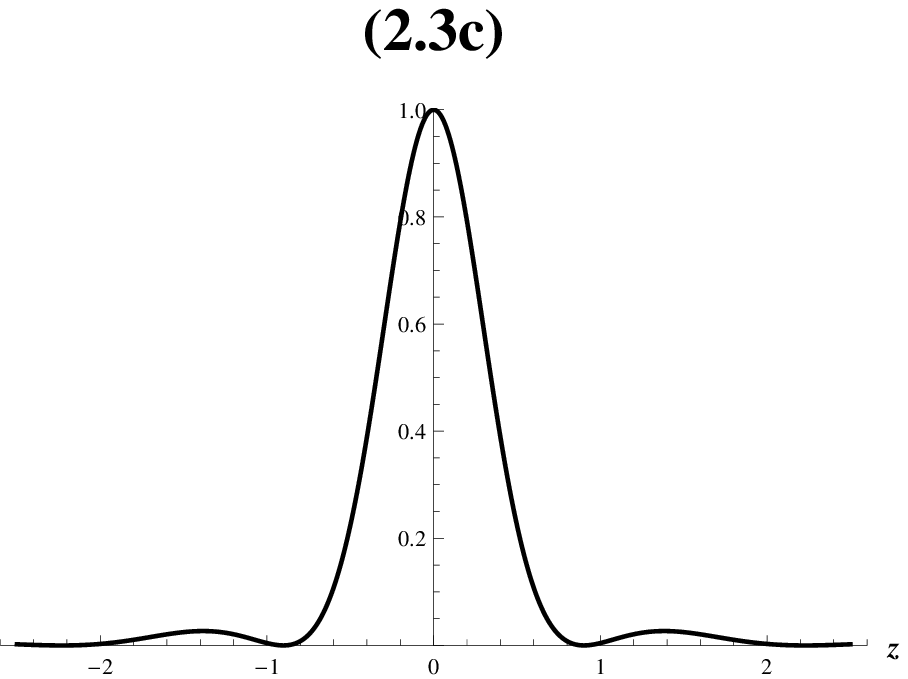}}
\scalebox{0.55}{\includegraphics{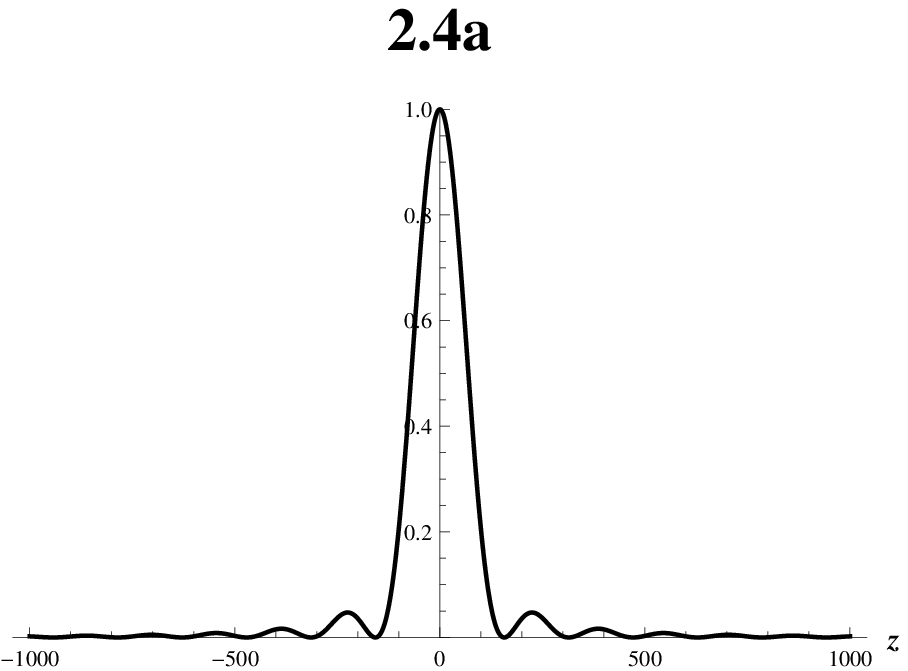}}\scalebox{0.55}{\includegraphics{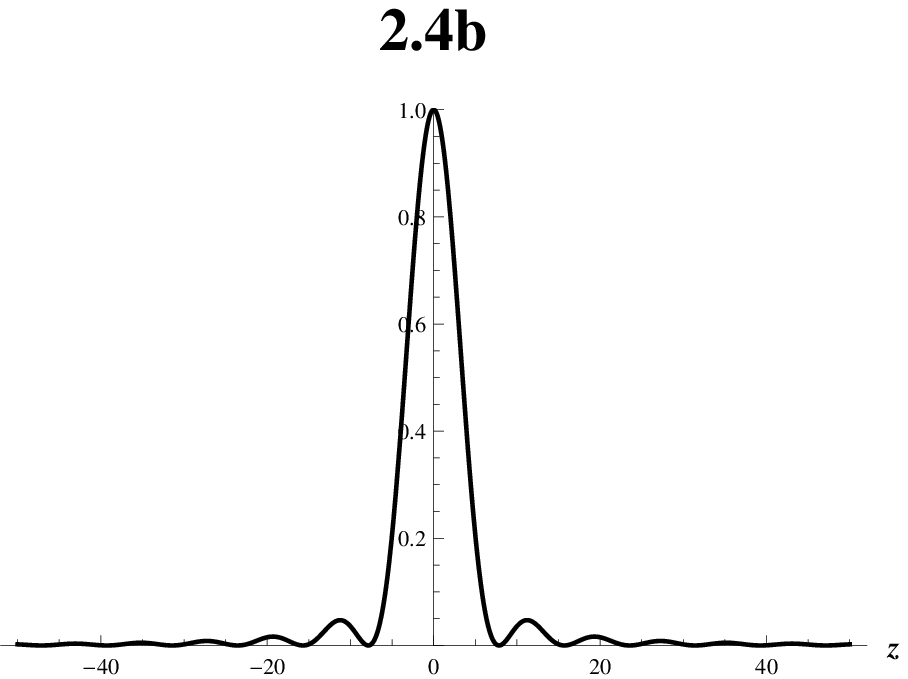}}\scalebox{0.55}{\includegraphics{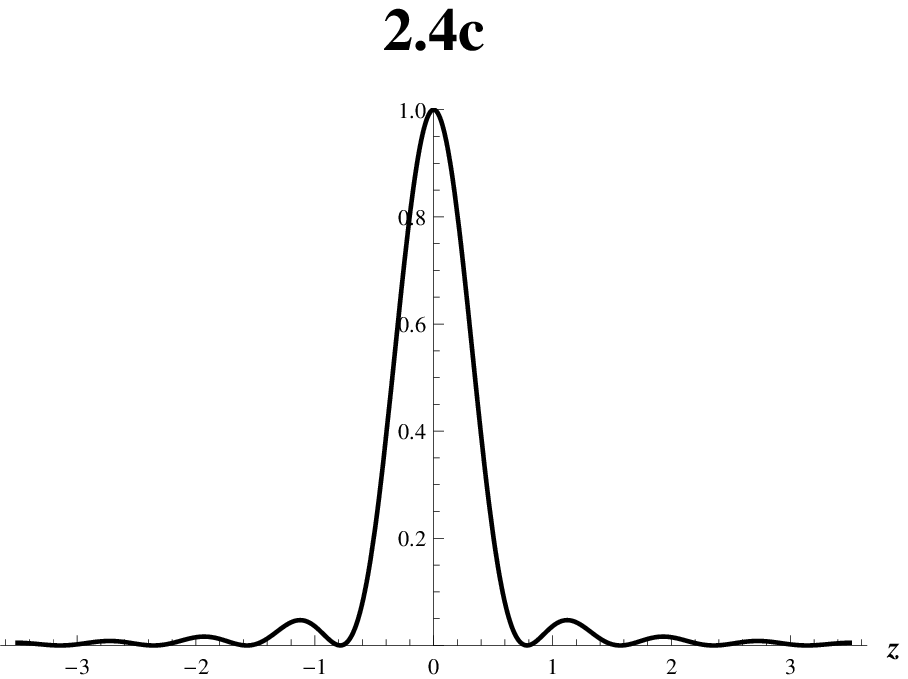}}
\caption{Semi-classical transition for the single-slit interference pattern.
We take $x=1,\ x_1=0,\ x_0=-1$, $a=0.01$ and $b=0.1$ in the units $\hbar=m=1$.
We represent the relative populations computed as the square modulus of the propagators (\ref{KRew})
respectively for the Dirichlet (Fig.2.1a-2.1c), Newmann (Fig.2.2a-2.2c) and free (Fig.2.3a-2.3c) boundary conditions
and also for the truncation approximation (Fig.2.4a-2.4c) by the Equation (\ref{Itr}),
with $t=1$ for the figures at the left (a), $t=0.05$ at the middle (b) and $t=0.005$ at the right (c).}
\end{center}
\end{figure}

\begin{figure}
\begin{center}
\scalebox{0.55}{\includegraphics{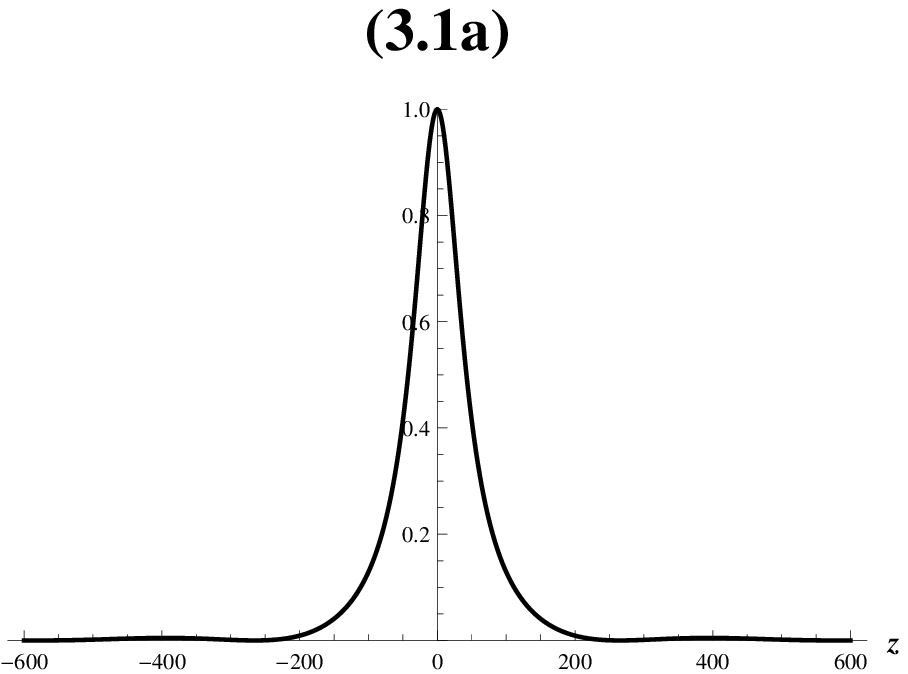}}\scalebox{0.55}{\includegraphics{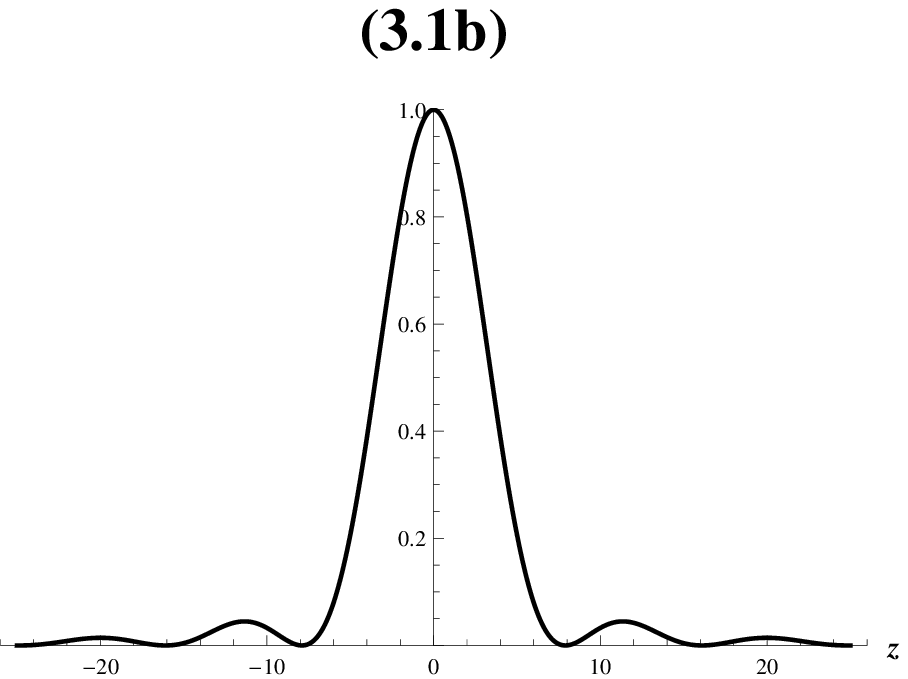}}\scalebox{0.55}{\includegraphics{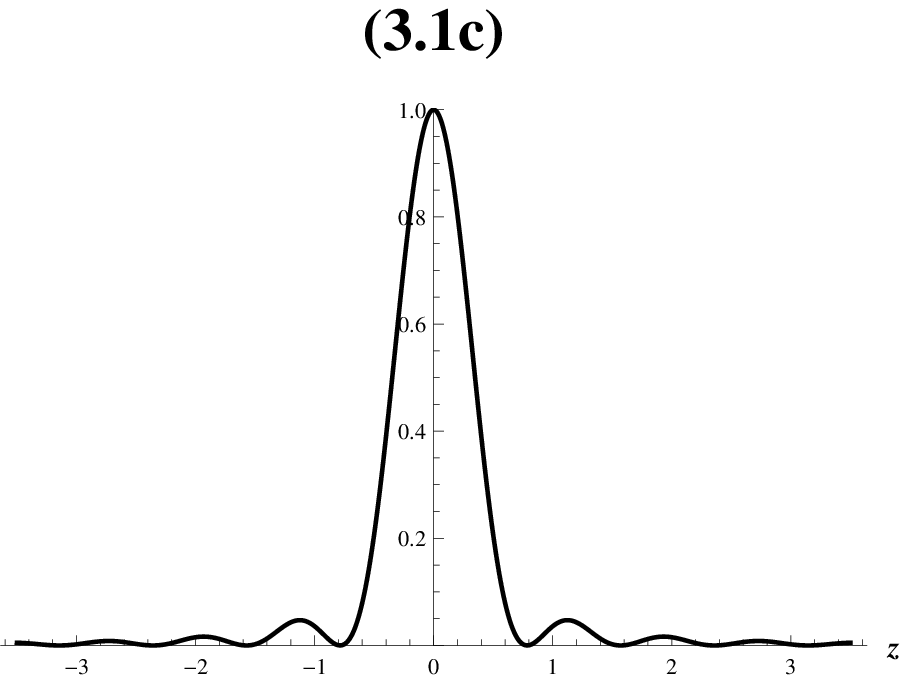}}
\scalebox{0.55}{\includegraphics{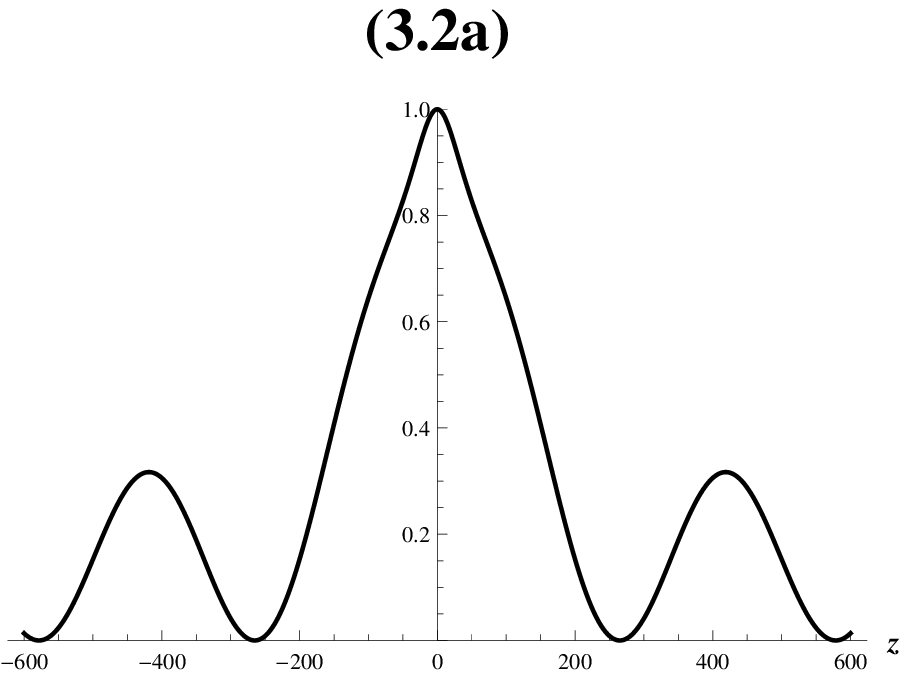}}\scalebox{0.55}{\includegraphics{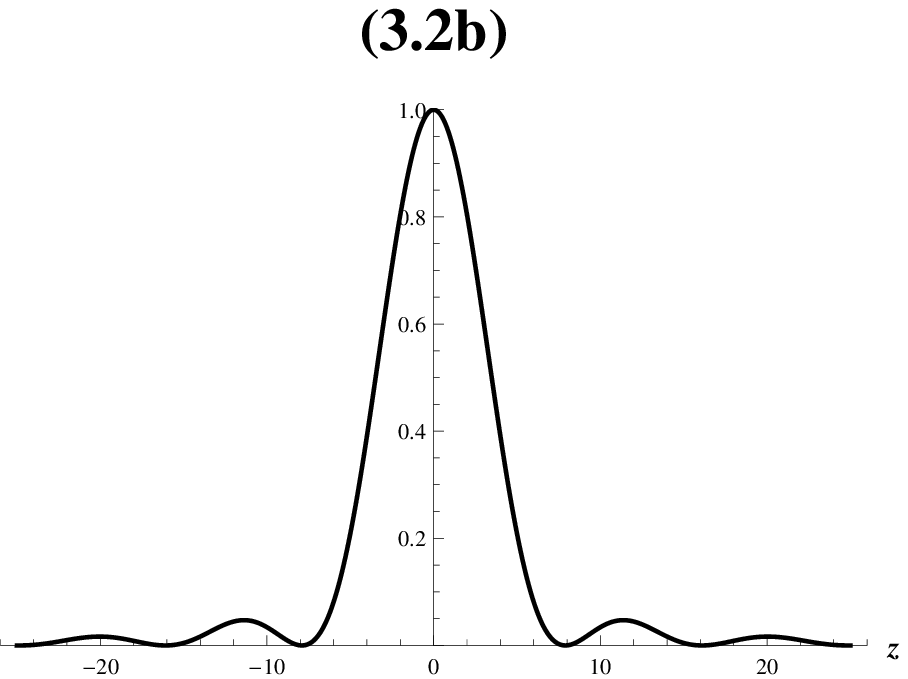}}\scalebox{0.55}{\includegraphics{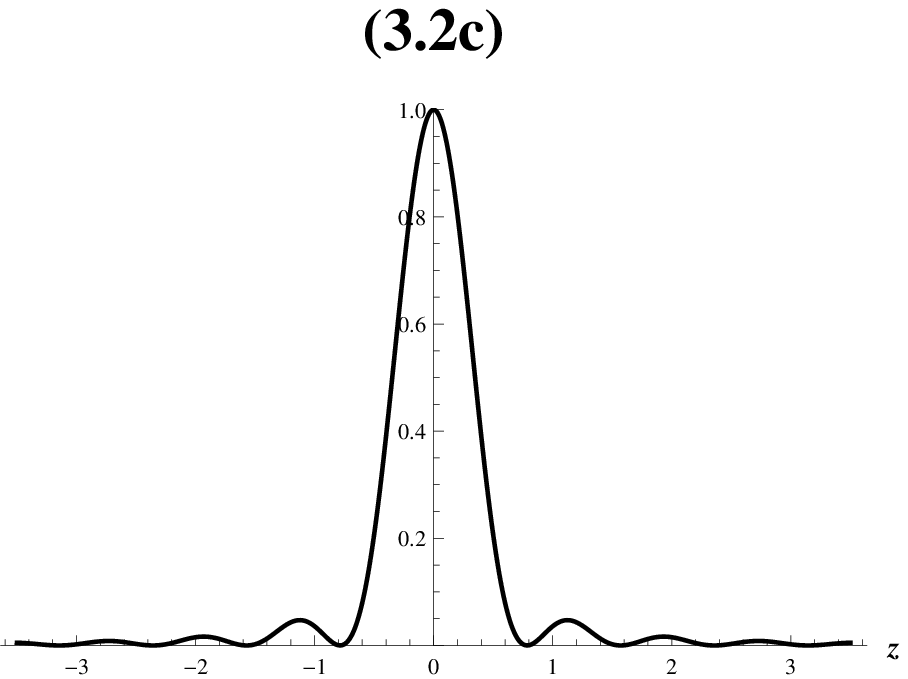}}
\scalebox{0.55}{\includegraphics{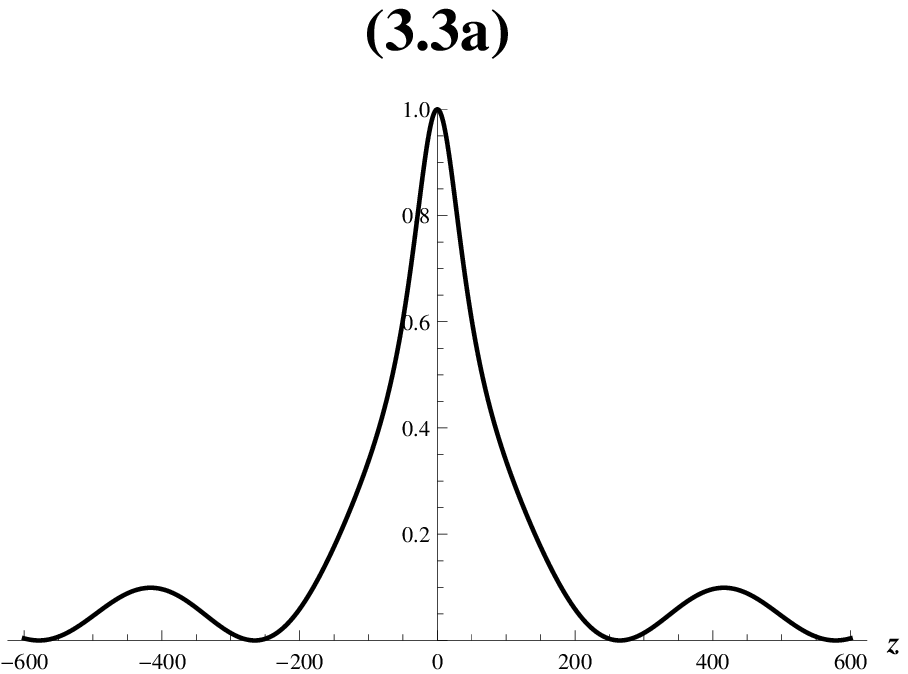}}\scalebox{0.55}{\includegraphics{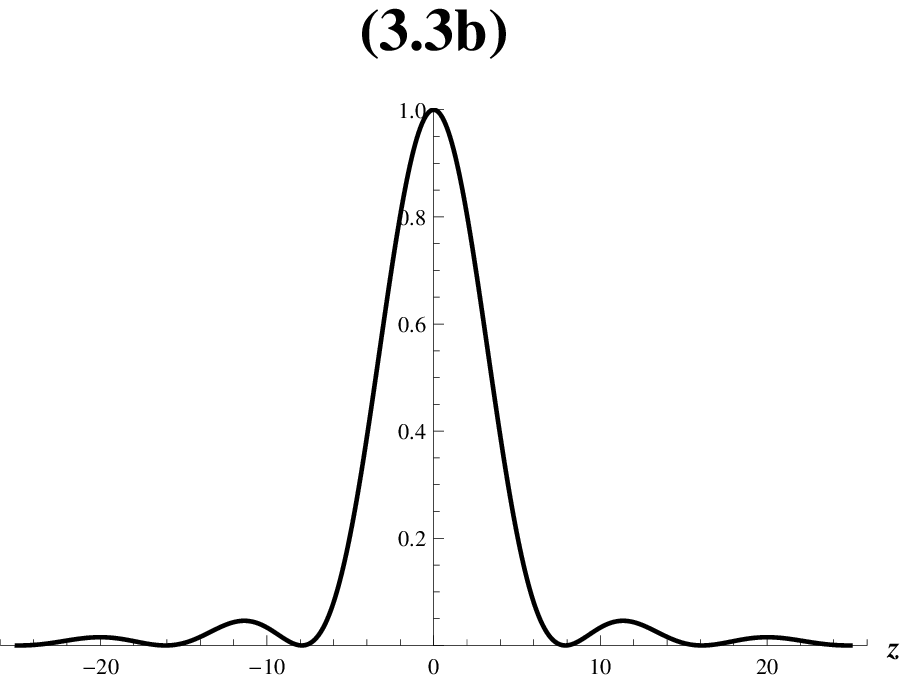}}\scalebox{0.55}{\includegraphics{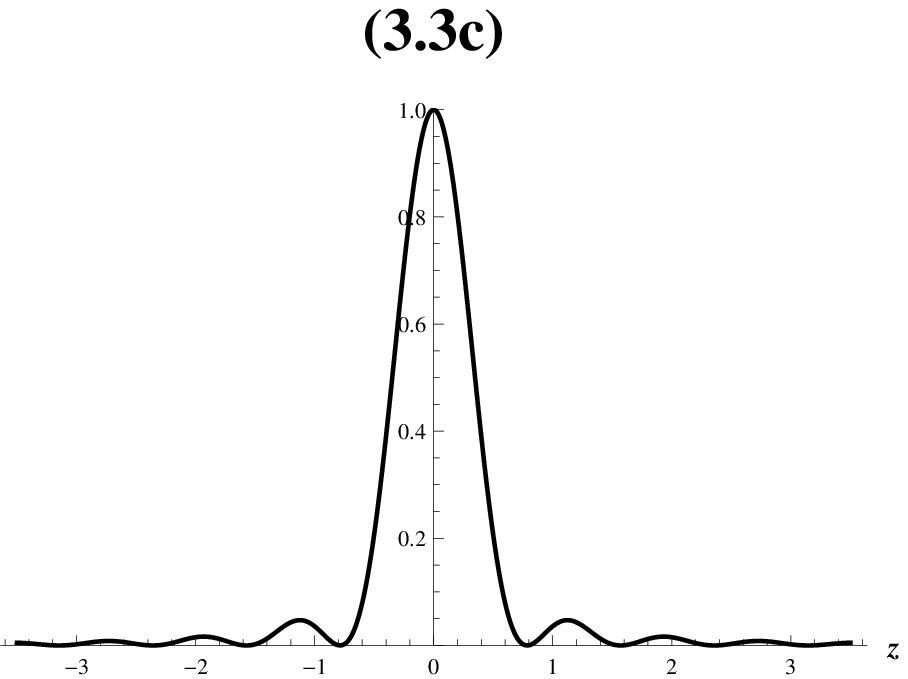}}
\scalebox{0.55}{\includegraphics{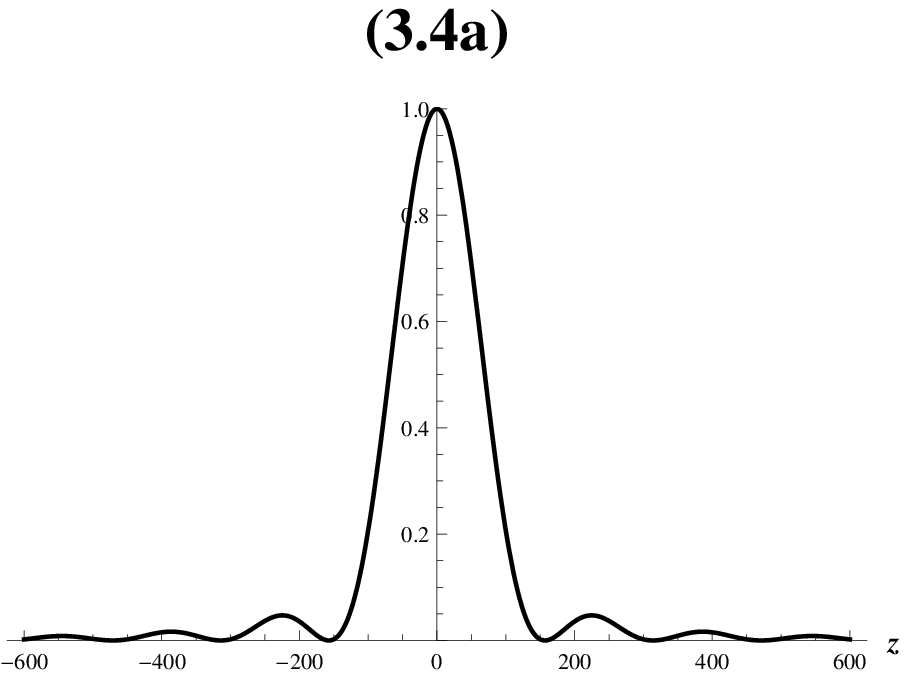}}\scalebox{0.55}{\includegraphics{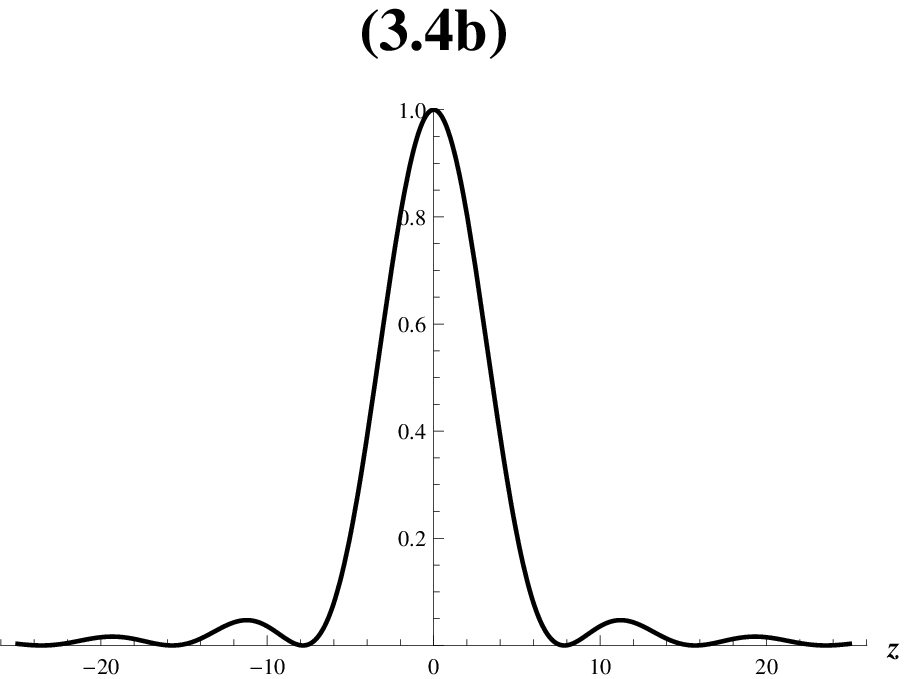}}\scalebox{0.55}{\includegraphics{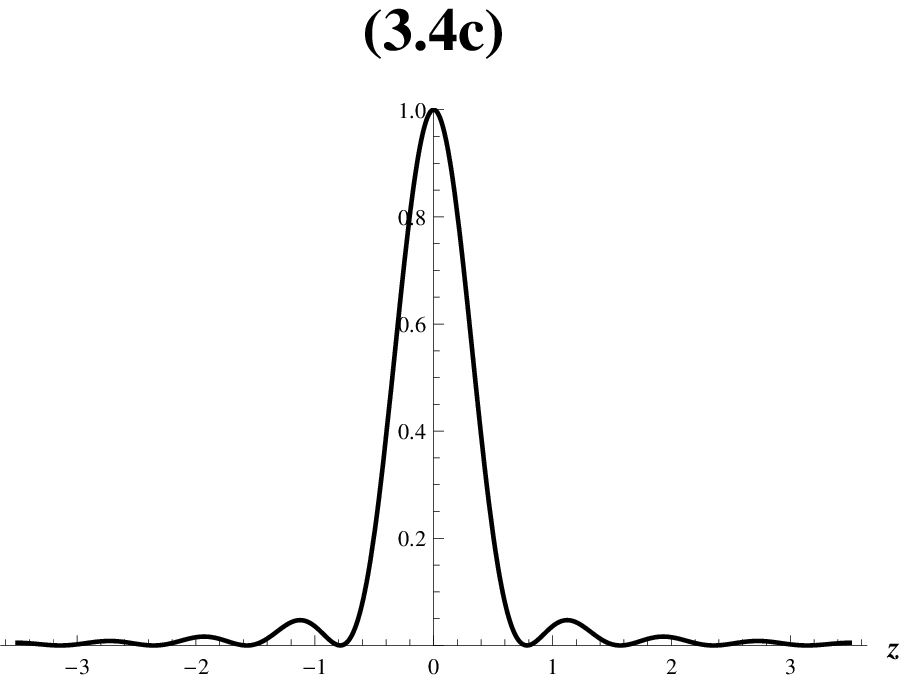}}
\caption{Truncation approximations for the single-slit interference pattern.
We take $x=50,\ x_1=0,\ x_0=-50$, $a=0.01$ and $b=0.1$ in the units $\hbar=m=1$.
We represent the relative populations computed as the square modulus of the propagators (\ref{KRew})
respectively for the Dirichlet (Fig.3.1a-3.1c), Newmann (Fig.3.2a-3.2c) and free (Fig.3.3a-3.3c) boundary conditions
and also for the truncation approximation (Fig.3.4a-3.4c) by the Equation (\ref{Itr}),
with $t=1$ for the figures at the left (a), $t=0.05$ at the middle (b) and $t=0.005$ at the right (c).}
\end{center}
\end{figure}

\begin{figure}
\begin{center}
\scalebox{0.75}{\includegraphics{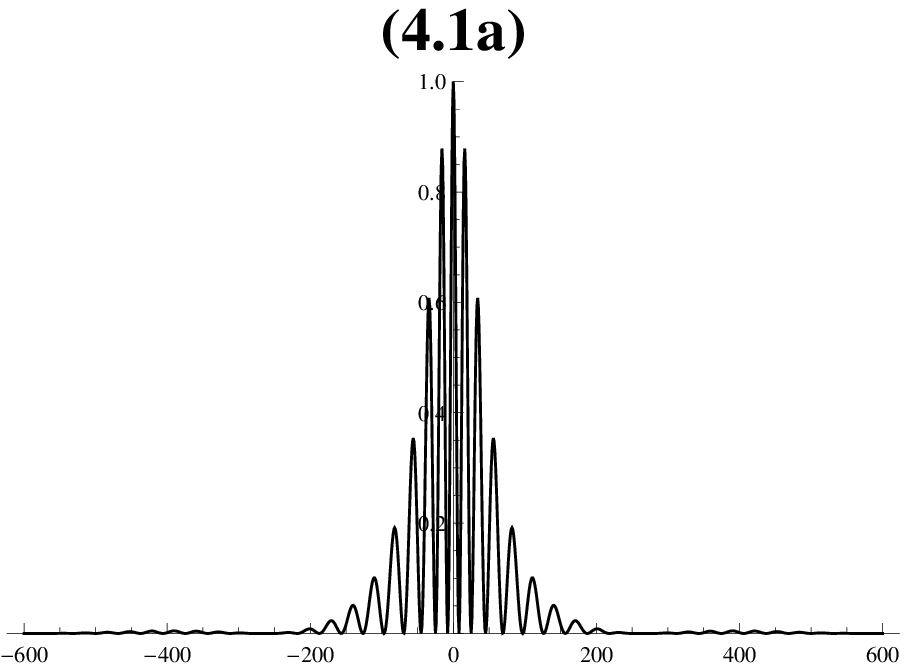}}\scalebox{0.75}{\includegraphics{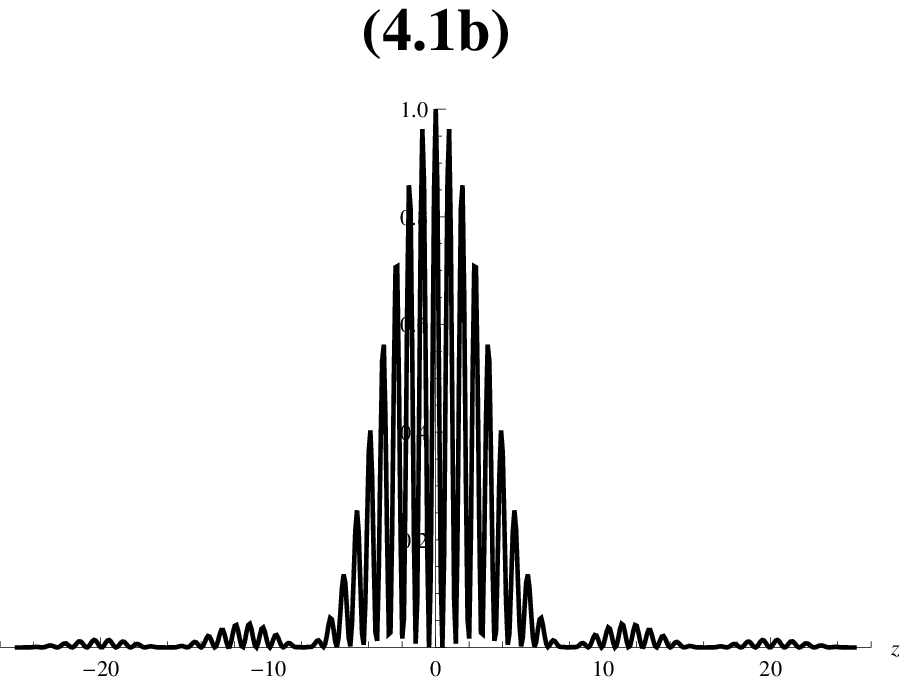}}
\scalebox{0.75}{\includegraphics{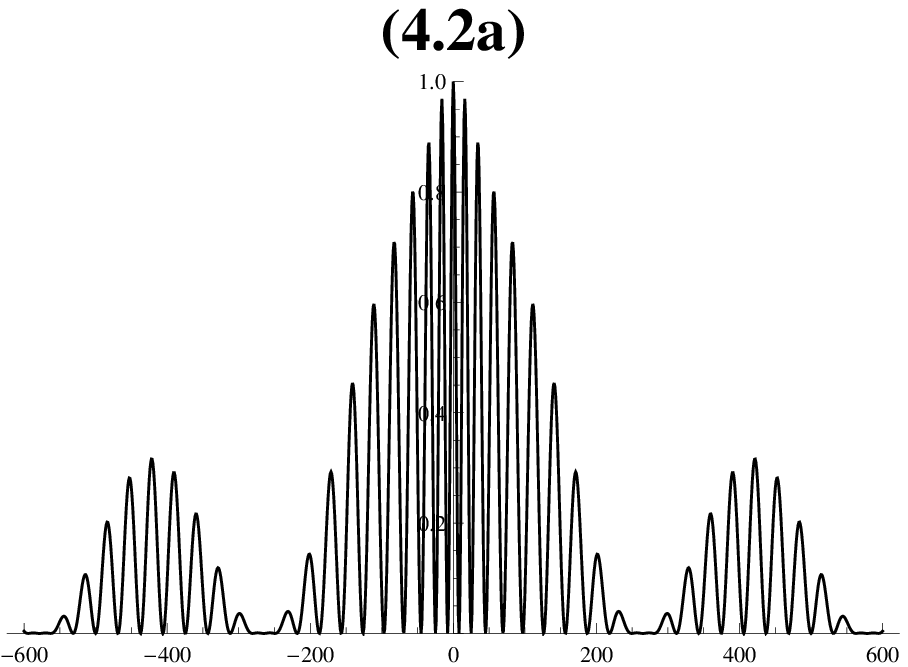}}\scalebox{0.75}{\includegraphics{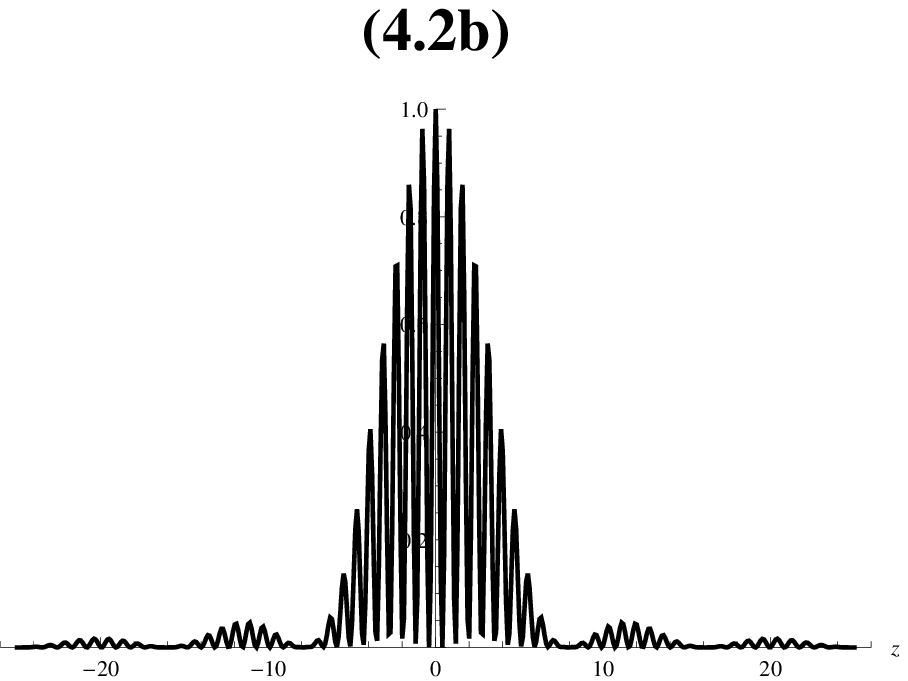}}
\scalebox{0.75}{\includegraphics{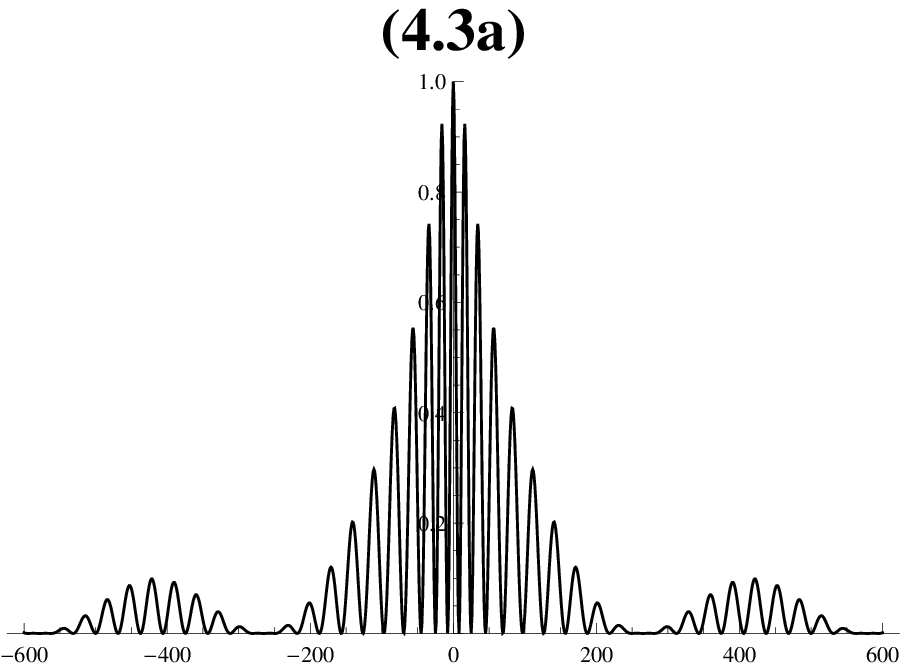}}\scalebox{0.75}{\includegraphics{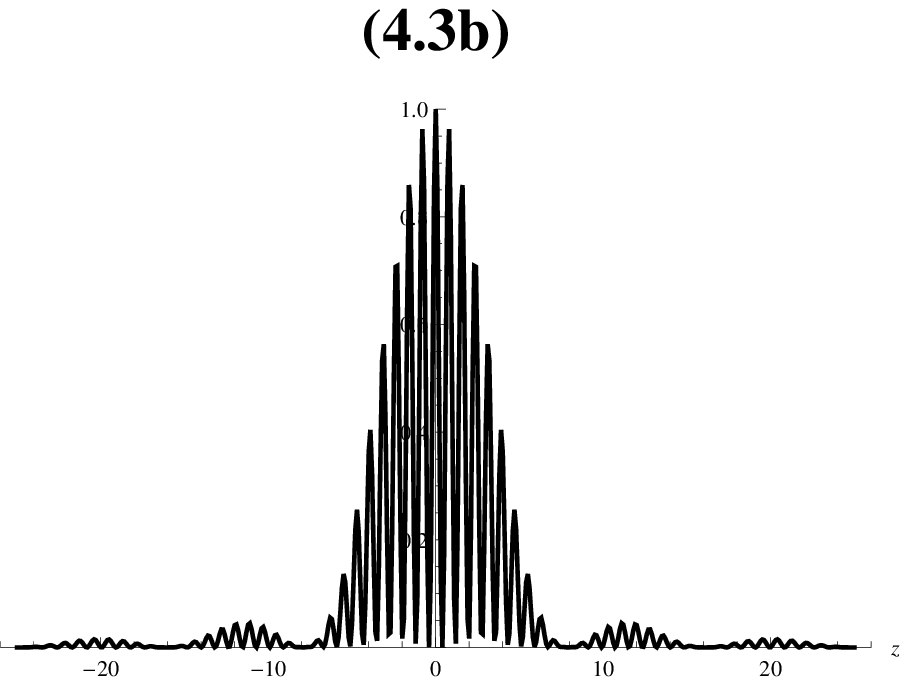}}
\caption{Double-slit interference patterns.
We take $x=50,\ x_1=0,\ x_0=-50$, $a=0.01$, $b=0.1$ and $d=0.1$, in the units $\hbar=m=1$.
We represent the relative populations computed as the square modulus of the propagators (\ref{Kdble})
respectively for the Dirichlet (Fig.4.1a-4.1b), Newmann (Fig.4.2a-4.2b) and free (Fig.4.3a-4.3b) boundary conditions
with $t=1$ for the figures at the left (a), $t=0.05$ at the right (b).}
\end{center}
\end{figure}

\end{document}